# On the construction of Tesla transformers

Period of oscillation and self-inductance of the coil.

(Zur construction von Teslatransformatoren. Schwingungsdauer und Selbstinduction von Drahtspulen)

By P Drude[1]

Annalen der Physik, 1902, vol. 314 (4[th] series, vol. 9),
Part I. issue 10, p293-339 + Plate 1, Part II. issue 11, p590-610.

Translated by David W Knight[2], and Robert S Weaver[3].
13[th] May. 2016.

Added by the translators:

## Table of Contents

---

1  Paul Karl Ludwig Drude, 1863 - 1906.
2  Ottery St Mary, Devon, England.  http://g3ynh.info/
   orcid.org/0000-0003-0499-3938
3  Saskatoon, Saskatchewan, Canada.  http://electronbunker.ca/
   orcid.org/0000-0002-7134-5095

## Translator's commentary

The following paper describes P K L Drude's 1902 investigation into the factors affecting the self-resonant behaviour of single-layer coils. This work, largely ignored by the English-speaking world in the second-half of the 20$^{th}$ Century, is particularly relevant to the construction of Tesla transformers, but it is also of general practical and theoretical interest.

The experimental method was that of exciting coils by means of an induction loop with a variable resonating capacitor, this circuit being energised by an induction coil and a Tesla transformer with both primary and secondary spark gaps. Resonance of the coil under investigation was detected by holding an electrodeless sodium-vapour discharge tube near to the coil. Wavelength calibration involved removing the coil and using the loop to excite a parallel-wire transmission line with a moveable shorting strap, resonance being detected by placing the discharge tube at the voltage anti-node.

The self-resonance period of a coil was found to increase with the dielectric constant of the core material; but this was less than proportional to the square root of the dielectric constant (as would be the case for immersion in a homogeneous medium). The dielectric effect of the core was also found to be greater as the height to diameter ratio was reduced, because of the increasing density of electric field lines on the inside of the coil. Hollow cylinders had less effect than solid cylinders. Wire insulation was also found to increase the self-resonance period, and the effect again increased as the height to diameter ratio was reduced.

Coils were characterised by means of a function $f$, which is defined as the ratio of the self-resonant half-wavelength to the wire length. Excluding dielectric effects, $f$ is primarily dependent on the coil height to diameter ratio ($h/2r$), its value being large when $h/2r$ is small. The effect of the pitch to wire-diameter ratio is relatively small, and the number of turns has little effect provided that there are more than 1. A graph of experimental values of $f$ vs. $h/2r$, for coils on ebonite cores ($\varepsilon = 2.79$) and for coreless coils, is given in Plate 1.

In accounting for the relationship between coil parameters and self-resonance, it was noted that when a current is induced in a disconnected long-thin coil, the current will be at its maximum in the middle region and zero at the ends. This causes electric charge to migrate towards the coil ends, inducing a potential difference. If the resulting charge displacement is considered to be localised on two squat cylinders located at the ends of the coil, the capacitance can be calculated in terms of spherical harmonics. The resulting calculated value of $f$ was within 5% of the observed over an $h/2r$ range from about 2.2 to 1.0.

Overtone resonances were also investigated, the node positions being located using the sodium-vapour discharge tube. Overtones are not harmonically related to the fundamental resonance. Drude argues that when a coil is oscillating at its first overtone, the fact that it does not behave as two separate coils is due to the magnetic coupling between the two halves.

When capacitance is added to a coil, such as by the addition of a conducting sphere at one end, the period of oscillation is increased, but never more than doubled. This effect can be quantified by considering the resulting shift in the voltage node.

In part II, the difficulty of calculating high-frequency inductance (i.e., the reduction due to skin effect and non-uniform current distribution) was overcome by placing fixed capacitances in parallel with coils and loops and making resonance measurements.

(DWK  June 2015)

## Notes on the translation

1) Translator's comments and additions are given in the text in [square brackets]. It is recommended that this document is read in conjunction with the original German papers, which are obtainable in a single pdf file from: https://archive.org/details/Drude1902Testlatrans

A short list of the obsolete or obscure German words and abbreviations that will be found in the original papers is given at the end of this document.

2) The source page number for the original German text is inserted into the text in square brackets, i.e., [p293] to [p339] and [p590] to [p610]. Note however that sentences that were split over two pages prior to translation are now placed entirely either before or after this number. Note also that source page numbers are not always in numerical order and may sometimes appear twice — this is because some tables have been moved to improve text flow and place them close to the text that refers to them.
    Footnotes are numbered sequentially in this document. For cross-referencing purposes, the original page number and footnote number is given at the beginning of the footnote, e.g, [297-3] indicates that the following text is a translation of footnote 3) on p297. In the absence of such a cross-reference, the footnote has been added in translation.

3) Drude used the cgs system of units. In some cases, such as capacitances and inductances in cm, the result in rational units has also been given in square brackets. To translate an inductance in cm to rational mks (SI), convert the length into metres and multiply by $\mu_0 / 4\pi = 10^{-7}$ H/m. This means that 1 cm ≡ 1 nH. To translate a capacitance in cm to SI, convert to metres and multiply by $4\pi\,\varepsilon_0 = 111.2650056$ pF/m. Thus 1 cm ≡ 1.112650056 pF. Frequencies in MHz are also given in some places using $f_0 = c / \lambda$, where c = 299 792 458 m/s.

4) Wiedemann's Annalen (Wied. Ann.) and Annalen der Physik (Ann. Phys or Ann. der Phys.) are the same journal, but with the volumes numbered in different ways (page numbers remain the same). References to Wied. Ann. are converted to references to Ann. Phys. by adding 236 to the volume number. The information needed for converting all early Annalen der Physik citations to the overall-series volume number is given at:
http://www.physik.uni-augsburg.de/annalen/history/history.html

# 3. On the construction of Tesla transformers
Period of oscillation and self-inductance of the coil
By P. Drude.

**Introduction**
The construction of Tesla transformers involves bringing a primary circuit formed by a coil of wire of few turns with applied end capacitance into electrical resonance with a coil of many turns without an applied end capacitance.  It will, especially in the construction of large and vigorously acting transformers, involve much time-consuming experimentation if the period of oscillation of the secondary coil and the inductance of the primary coil cannot be calculated in advance.  This matter will be addressed in the following article.  A rational approach to the dimensioning of Tesla transformers, and to their theory, will be covered in a later paper[4].  — The knowledge of the natural period of coils can also be applied to the construction of the important newer devices for wireless telegraphy, although one should be careful that the electrical conditions may be essentially different if the coil does not have free ends but has capacitance or straight wires connected.  The resulting changes can be estimated theoretically, but the period of oscillation of the free-ended coil must first be known.

   Note that here we discuss only coils of wire in a given winding sense, i.e., with large self-inductance, as they are primarily the Tesla coils of importance.  — For the purpose of wireless telegraphy, and also for laboratory experiments with Tesla coils, wire coils with different winding sense, i.e., smaller self-inductance, are sometimes useful.  To keep this document reasonably small, such coils are excluded from this discussion.

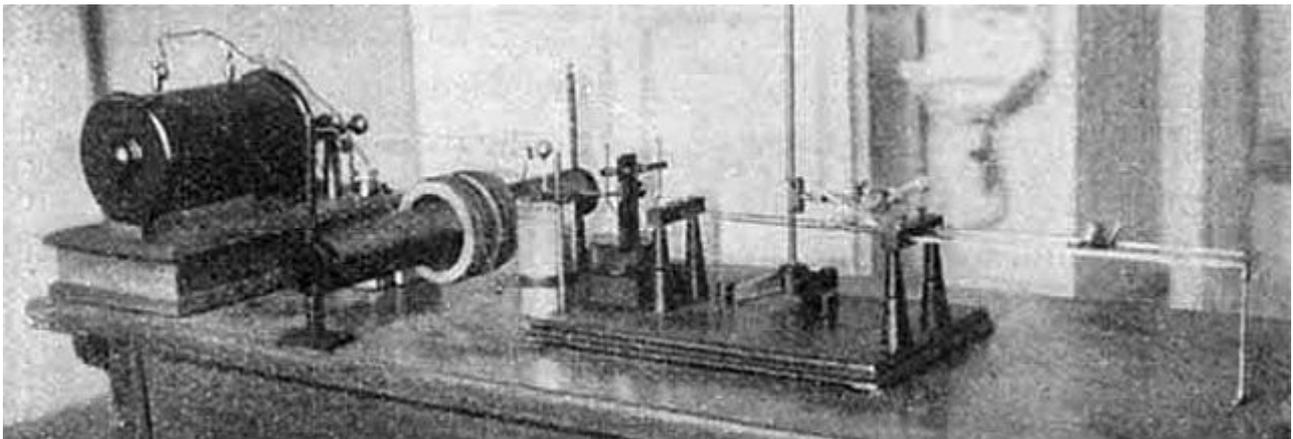

[A photograph of apparatus similar to that used in this paper (the induction coil on the left is almost certainly the same one).  This picture is from '**Zur Messung der Dielektricitätsconstante vermittelst elektrischer Drahtwellen**' (Measuring the dielectric const. by means of electric wire waves (i.e. standing waves) ), P. Drude, Ann. Phys. 313(6) (4[th] series vol. 8). p336-347. 1902.  This is Fig. 2 from page 340.]

---

[4] **Rationelle konstruction von Teslatransformatoren**, P Drude.  Ann. Phys. Vol. 321(1), 1905, p116-133

# I. Oscillation period of wire coils

## 1. Experimental method

The experimental method was that the coil to be examined, *S*, was excited inductively by the electrical oscillation of an exciter *E* (Fig. 1), which consisted of two semi-circularly curved 3 mm thick copper wires that spanned a circular area of 21 cm diameter. The exciter wires were held by two thick slotted ebonite [hard rubber] supports, *H*. At one end they were bent down so that they were immersed in a glass bowl filled with petroleum. This end had small brass balls of 0.5 cm diameter, the separation (about 0.25 mm) easily varied by a shift of the ebonite support *H*. The excitation spark between the brass balls took place under Petroleum. They were connected to the ends of the secondary coil, *T*, of a Tesla transformer[5], which was fed by an Induction coil, *J*, with striking distance of 30 cm, having a rotary mercury breaker. *Z* is the zinc spark gap for wave excitation in the primary coil of the Tesla transformer, *L* is the Leyden jar of its primary circuit.

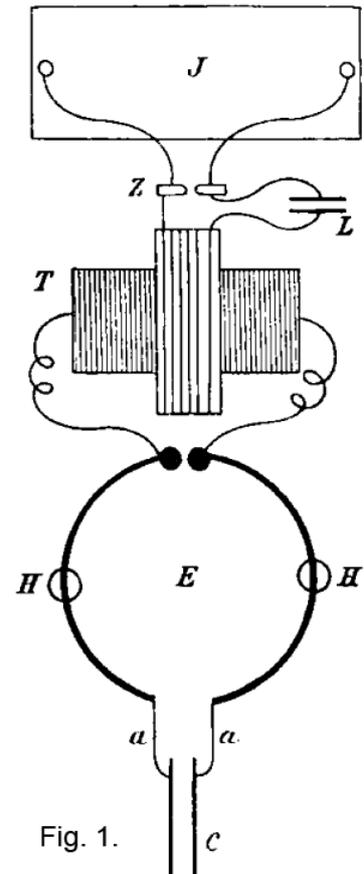

The other ends of the field wires lead to two 9 cm long, 0.5 mm thick copper wires, *a*, *a*, which connect to a petroleum-immersed circular plate capacitor, *C* ( Fig. 2).

- [p295] -

A petroleum bath *P* for *C* is extremely convenient, because it allows the distance of the plates to be reduced to 1 mm without the occurrence of spark or corona discharge. The capacitance can therefore be varied over a much wider range than when *C* has an air environment. The plates of *C* were 10 cm in diameter, their separation could be up to 5 cm. They were attached, with vertical ebonite supports *e*, *e*, to two horizontal arms, *h* , *h*, one of which was moveable and had a scale for measuring parallel displacement. The distance of the ends of the excitation wires, from which the thin copper wires were angled, was 5 cm.

The coil to be tested, *S*, was placed vertically in the centre of the excitation loop on wooden blocks. Depending on the circumstances, the distance was 5 cm - 30 cm from the excitation plane, and a vacuum tube was placed at the end. The tube was placed in the open, 1 cm - 2 cm from the end of the coil. Electrodeless tubes by the glass blower Kramer in Freiburg in particular are highly

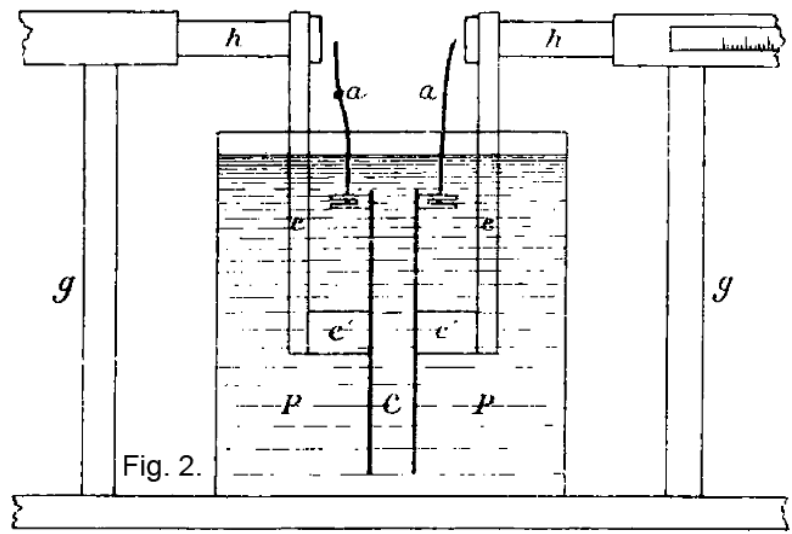

Fig. 1.

Fig. 2.

---

5 [294-1] By inserting a Tesla transformer between inductor and exciting spark gap , the intensity of the electric waves is greatly increased. The dimensions of the Tesla transformer , which are quite irrelevant if it is sufficiently vigorous, were as follows: Secondary coil 100 turns of 1 mm thick (with insulation 2 mm thick) copper wire on a wooden cylinder of diameter of 9 cm and 20 cm height. The Leyden jar *L* was 11.5 cm in diameter, with 19 cm overlap height, and 5 mm thick glass.

recommended. These have a thin layer of electrolytically deposited sodium.

When the induction coil is operated in a darkened room, and there is a spark discharge between the balls of the exciter, the vacuum tube will not generally be lit. Only at a certain separation of the plates of capacitor $C$ will it light up.

- [p296] -

This separation corresponds to the resonance[6] between the wire coil and excitation circuit.

This resonance position of the capacitor $C$ is determined by adjusting a horizontal arm on which a plate of $C$ is fixed. The resonance becomes sharper as the inductive excitation (magnetic coupling) of the coil by the exciter is reduced, i.e., the higher it is above its level, provided that the resulting illumination of the vacuum tube is not too weak. Weak magnetic coupling between the two systems is however necessary, because otherwise (except for the attenuation of the excitation vibrations, see footnote [296-1] ) maximum excitation of the coil will not exactly occur at resonance because of the retro-action of the coil on the exciter circuit. This pulling effect however, was not generally a problem, because usually the diameter of the test coil (2 cm - 3 cm) was much smaller than the diameter of the excitation loop (21 cm), so that the number of flux linkages was small. In any case however, the distance between the coil and the exciter was always large, and the spark gap was kept small to achieve preferably weak rather than strong illumination of the vacuum tubes. This is because, if the tube glows strongly, the conductivity of the gas is significantly increased, and if this is applied at one end of the coil, its period of oscillation is slightly reduced in comparison to a coil with two free ends

There was thus obtained, in a coil of 30 cm length and 1.7 cm diameter, which consisted of 100 turns of 1 mm thick bare copper wire, a resonance distance $d$ between the plates of the capacitor $C$, $d = 18.7$ mm, that is, $\lambda/2 = 286$ cm, when the tubes glowed strongly. Contrast that to $d = 21.0$ mm, that is, $\lambda/2 = 277$ cm, if the tubes glowed weakly.

- [p297] -

The influence of capacitance increase caused by the glowing tubes is reduced as the self-resonance wavelength $\lambda$ of the coil gets longer.

The resonance positions of the capacitor $C$ were adjusted several times (usually 6 times) and the distance of the capacitor-plates was measured using a vernier scale with 0.1 mm accuracy[7]. An experiment with a particular coil was terminated only after settings were found such that a small change in the intensity of the inductive excitation, or in the distance[8] of the vacuum tube from the coil end, and thus the light intensity of the vacuum tube, had no noticeable influence in the natural oscillation period of the coil[9]. The coil ends were often held by small wire pins. Confirmatory tests showed that the same settings were obtained when the coil ends were cemented with sealing wax, or held by notches in the coil core or by twine.

The oscillation periods associated with each spacing of the capacitor plates $C$ were obtained by

---

6 [296-1] This does not strictly apply if the attenuation of the excitation vibrations is very significant (see: '**Periode für welche die Amplitude einer erzwungenen Schwingung ein maximum wird**' [Period for which the amplitude of forced oscillation is a maximum], M. Wien , Ann. Phys 294(8) 1896 (Wied. Ann. 58) p725-728). It is so small here however that it can be neglected. As calculated from the attenuation of the excitation without applied end capacitance $\gamma = 0.15$ (see '**Theorie stehender electrischer Drahtwellen**' [Theory of electric wire standing waves], P Drude , Ann. Phys. 296(1) 1897 (Wied. Ann. 60), p1-46, see p17) $k = 0.05$ n according to Wein. Now, if the attenuation with applied end capacitor is large, it will still only gain influence (0.5 %), when three times as large, i.e., when $\gamma = 0.45$, and this is certainly not the case.
7 [297-1] It was even possible to estimate to 0.02 mm by using a magnifying glass.
8 [297-2] In the case of the great intensity of electrical oscillations this distance could be 3 cm; e.g., with a 16cm long coil, this was possible even if the lower coil end was 15 cm above the exciter plane and the upper end was 31 cm above it.
9 [297-3] Although instead of lighting a vacuum tube, a small spark gap at the end of the coil could be used as a wave indicator and gave the same resonant distances $d$ of the capacitor plates. Spark gaps however are not such sensitive indicators as vacuum tubes.

calibrating the apparatus in the following manner:

After measuring a coil, *S*, a 7 m long transmission-line, *D*, made of two bare 1 mm diameter copper wires stretched taut was arranged 15 cm above the level of the excitation wires (Fig. 3.).

- [p298] -

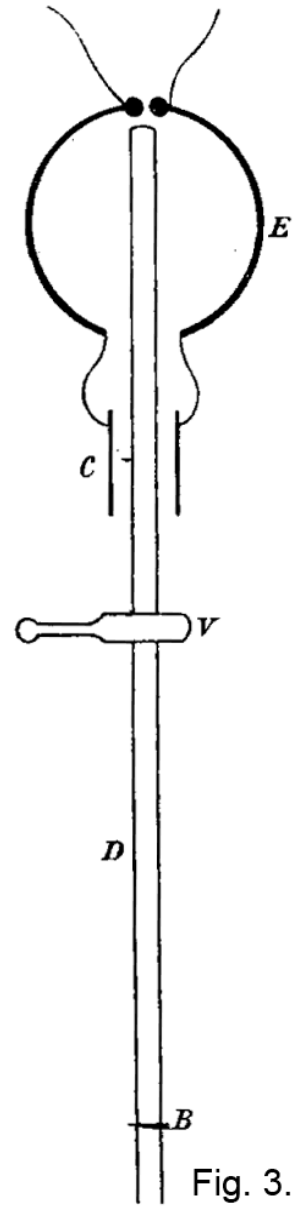

Fig. 3.

The wire spacing was 2.7 cm. They were shorted close to the spark gap (the beginning of the line *D* ). At the other end they were shorted by a sliding metal strap *B*. The shorting strap was moved by hand (in a darkened environment) so that a vacuum tube *V* placed approximately half way between *B* and the beginning of the line glowed at maximum brightness. This occurs when the resonance[10] of the line *D* is consistent with the oscillations of the exciter *E*. Each spacing of the plates of the capacitor *C* therefore corresponds to a particular resonance position of *B*. Because of the weak magnetic coupling between *E* and *D*, these resonance settings are very sharp (0.25% to 0.5% of the distance from the beginning of the line *D* to the strap *B* ). The half-wavelength of the electrical oscillation is equal to the distance of the shorting strap *B* from the beginning of the line, increased[11] by the length of the shorting straps; plus a small addition due to capacitance of the glowing vacuum tube. The latter was noticeable here because, when observing long waves, the vacuum tube had to be removed so far (3 m) that the faint glow would no longer be perceived. Both corrections can be determined exactly (at shorter wavelengths), by leaving *V* where it is and moving *B* further back to the next resonant position.

- [p299] -

The distance between the first and second resonance positions of *B* is exactly one half-wavelength. The correction so obtained was an addition of 9 cm[12], although it depends to some extent on the actual wavelength. This latter variation however is so small that it was within the observational error (0.25 %) and could be neglected. This calibration of the exciter was carried out always immediately before and after an observation of a coil *S*. The calibration results changed significantly only when the plate-capacitor *C* was taken apart and reassembled. The following table contains the results. *d* is the spacing of the capacitor plates expressed in millimetres, $\lambda/2$ the corresponding half-wavelength of the exciter oscillation in centimetres .

---

10 [298-1] Due to the large distance between *E* and *D* and because of the small relative distance between the two wires *D*, the magnetic coupling between exciter *E* and line *D* is so weak that a reaction from *D* to *E* is not noticeable. Thus the position of *B* for which *V* glows most brightly really corresponds to the resonance. This was proved by the fact that the position of *B* is not dependent on the distance between *D* and *E*.

11 [298-2] See '**Theorie stehender electrischer Drahtwellen** [Theory of electric wire standing waves], P. Drude, Ann. der. Phys, Vol. 296(1) (Wied. Ann. 60), 1897, p1-46, see p14.

12 [299-1] 3 cm was omitted from this correction due to the proximity of the wooden measuring rod (2 cm) above which the parallel wires were strung. Because this distance of 2 cm was increased to 6.5 cm, so there was only 6 cm additional correction , instead of 9 cm. Therefore, the shorting straps contribute 3 cm, the capacitance of the glowing vacuum tube another 3 cm, and the proximity of the wooden measuring rod 3 cm. The additional correction (9 cm), which is always applied in the following, gives the correct wavelength in free air, because the proximity of wood for the rear parts of the secondary line was avoided.

Dependence of the wavelength λ of the exciter on the distance d of the capacitor plates.

| d / mm | 3 | 5 | 7 | 9 | 11 | 13 | 15 | 17 | 19 | 22 | 26 | 31 | 39 | 50 |
|---|---|---|---|---|---|---|---|---|---|---|---|---|---|---|
| λ/2 / cm | 585 | 467 | 408 | 369 | 343 | 324 | 309 | 296 | 287 | 274 | 262 | 251 | 237 | 226 |
| $f_0$ / MHz | 25.6 | 32.1 | 36.7 | 40.6 | 43.7 | 46.3 | 48.5 | 50.6 | 52.2 | 54.7 | 57.2 | 59.7 | 63.2 | 66.3 |

It was difficult to measure the absolute value of *d* accurately[13], so there is an uncertainty of up to 0.1 mm in *d*.

- [p300] -

This is not obvious because the plate spacing *d*, which was read on the scale on the horizontal arm of one of the plates of the capacitor *C*, was the same between calibration and observation of a coil, provided that the capacitor *C* had not been taken apart, and provided that only a short time[14] had elapsed between the two observations. As said above, the capacitor plates were screwed to two vertical ebonite holders *e, e*, and these in turn were screwed to two horizontal metal arms *h, h*, which rested in sliding guides on insulating glass pillars *g, g* (Fig. 2).

- [p301] -

During the resonance setting for a coil under observation *S*, the sliding horizontal arm (adjusted by micrometer screw) was touched by a hand, i.e., shunted to earth. This caused the capacitance of the capacitor *C* to be somewhat increased, as if the two horizontal arms were insulated. However, this was only noticeable when the distance of the capacitor plates *C* was large (approaching 5 cm), i.e., when the capacitance was small. Whether this was significant or not, could be easily detected during the calibration procedure; by noting whether the resonance positions for the shorting strap *B* were different if the horizontal arms supporting the plates of the capacitor *C* were insulated or were

---

13 [299-2] To the nearest 0.1 mm, the absolute values of *d* are about right. For sufficiently large capacitance of *C*, i.e., sufficiently small *d*, λ/2 becomes proportional to √C, i.e., inversely proportional to √d. For *d* = 3, 5, 7 mm we get (λ/2).√d as 1012, 1043, 1078 ; this inconsistency is because the approximation formula $C = r^2 / 4d$, where *r* is the radius of the capacitor plates, is used instead of the more accurate formula (see: e.g., B F Kohlrausch, Leitfad. d. prakt. Phys. 8$^{th}$ ed. p409):

$C = (r^2/4d) + r / 4π ( \log_e\{ 16 π r [d+δ] / d^2 \} -1 + [ δ/d ] \log_e\{ [d+δ] / δ \} )$

where δ is the plate thickness. The dimensions were δ = 1 mm, *r* = 5 cm. The wavelength is yet to be multiplied by √ε, where ε is the dielectric constant of petroleum. It was found that √ε = 1.41, this being the ratio of the exciter wavelengths when compared with using the capacitor in air. If we calculate *C* using the above formula and multiply by 1.41 we obtain:

| d | 3 | 5 | 7 | 9 | 11 | 13 | 15 | 17 | 19 | 22 | 26 | 31 | 39 |
|---|---|---|---|---|---|---|---|---|---|---|---|---|---|
| (λ/2)/√C | 85.6 | 85.9 | 86.5 | 87.2 | 88.3 | 89.4 | 90.7 | 90.5 | 91.7 | 92.4 | 93.9 | 96.1 | 98.5 |

i.e., it is actually the case that λ/2 ~ √C, and the deviation for larger values of *d* is considerable because the formula for *C* is then still too imprecise. The regular increase of the quantity (λ/2):√C however supports the reliability of the observations. If we take the value 85.6 for *d* = 3 mm as reasonable, we can use the formula $λ = 2π\sqrt{LC}$, where *L* is the self-inductance of the exciter loop. This gives the value *L* = 744 cm [= 744 nH]. According to M Wein, (Ann. Phys. 289 (Wied. Ann. 53). p931. 1894), for a wire of length ℓ and thickness 2ρ, spanning a circular area of radius *r* ; we have
$L = 2ℓ ( \log_e\{8r/ρ\} - 2)$. Here we have ℓ = 2.32 cm, as each half of the exciter loop was 32 cm long, 2*r* = 21 cm, ρ = 1.5 mm. To this value of *L*, the self-inductance of the two 0.5 mm thick, 9 cm long wires *a* is still to be added. For two parallel wires of length ℓ', thickness 2ρ', whose relative distance is *d* ', we get (see P. Drude, Physik des Aethers, p364) $L' = 4ℓ' \log_e(d'/ρ')$. Here we have ℓ' = 9, ρ' = 0.025, *d* ' = 5. Thus the sum
*L* = 554 + 188 = 742 cm. This is in precise compliance with the value of *L* (744 cm) obtained from λ/2 and *C*, but with some error as the wires could not be run exactly parallel because of their connection to the capacitor plates.

14 [300-1] If the capacitor *C* remains for several days in petroleum, then the 7 mm thick, 15 mm wide, 12 cm long ebonite arms carrying the plates of the capacitor bend noticeably. Within the observation time between two measurements however (2 hours), such deflection is not noticeable.

earthed. At a plate distance $d = 4.8$ cm the following were obtained: $\lambda/2 = 224.5$ cm if both horizontal arms were insulated; $\lambda/2 = 225.0$ if one horizontal arm was grounded; $\lambda/2 = 227.0$ if both horizontal arms were grounded.

- [p302] -

For smaller plate distances $d$, the changes of $\lambda/2$ by grounding the horizontal arms were small or imperceptible. Since, during the observations of the coils, only one horizontal arm was grounded, and the distance $d$ was almost always smaller than 3 cm, the resulting capacitance change was negligible assuming that measurement accuracy of 0.25% is acceptable. — In contrast, such a capacitance effect was very noticeable when the ebonite holders $e$, $e$ (Fig. 2.) were replaced by metal strips, while $e'$, $e'$ (Fig. 2) were made of ebonite. In that case; at $d = 4.8$cm: $\lambda/2 = 235$ cm with the horizontal arms isolated; $\lambda/2 = 245.5$ cm with one horizontal arm grounded; $\lambda/2 = 275.5$ cm with both metal arms grounded. Usually the vertical brass holders were replaced by ebonite holders.

The results of calibration were plotted as a graph, and the corresponding $\lambda/2$ at any $d$ taken from it. Fig. 4 is a scaled reproduction of that curve.

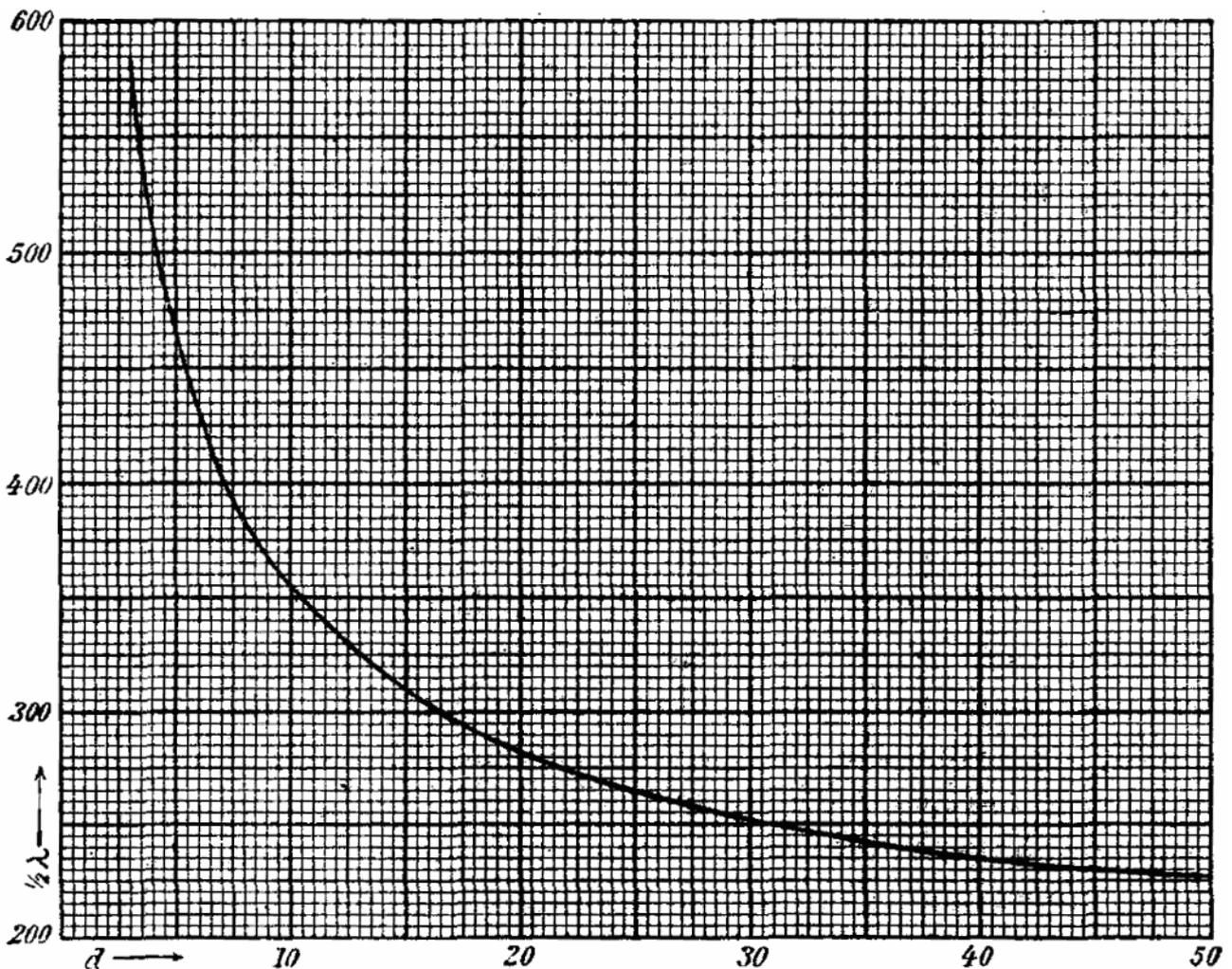

Fig. 4. [p301]

A second calibration method of wave exciters in the range $\lambda/2 = 6$ m to $\lambda/2 = 12$ m is discussed later (in Part II) [see p598 - 599].

## 2. Transfer to large coils of the results obtained on small coils

Since the natural resonances of the test coils could not exceed the corresponding half-wavelength of $\lambda/2 = 6$ m or 12 m, depending on whether the first or the second exciter calibration method was used, only relatively small coils were examined. It is possible however to transfer the results obtained for them to larger, geometrically similar coils, because a consequence of Maxwell's electromagnetic-field equations is that *the natural oscillation periods of geometrically similar systems scale exactly in proportion to the physical dimensions*[15].

## 3. Effect of the nature of the coil core and and its area on the self-resonance period

By winding a particular type of wire in exactly the same geometrical arrangement on cylinders of different materials, *the self-resonance period of the coil increases with the dielectric constant of the coil core*; but the rate is slower than proportional to the square-root of its dielectric constant.

- [p303] -

This is easy to understand, since the period of the coil must increase proportionally with the square-root of dielectric constant of the environment when the coil is in an infinite homogeneous medium. The fundamental electric oscillation now takes place in such a way that, in the middle of the wire length the oscillating current has maximum amplitude of vibration, in contrast to the potential at the ends. The ends therefore have periodically-varying positive and negative free electric charge. Between the coil ends therefore, there are induced electric field-lines, mostly outside of the coil, but to a small extent also inside the coil; and in the latter case the more so the shorter the coil is relative to its diameter. If now the dielectric constant increases in the interior of the coil, it must increase between the coil ends, thus increasing the self-resonance period of the coil when the dielectric constant of the core is large; specifically, because of the increasing density of electric-field lines in the interior of the coil, the capacitance increases as the coil becomes shorter relative to its diameter. *Coils on hollow cylindrical insulating material therefore have shorter self-resonance periods than coils on solid cylinders*; the more so, of course, the thinner the [wall of the] hollow cylinder is.
    *If the coil is immersed in a bath of liquid insulator* (petroleum) *instead of air, the self-resonance period will increase in consequence* (because of the electric field lines outside the coil).

**Some examples to illustrate this proposition**:
    A coil of 100 turns of 1 mm thick bare copper wire, of 15 mm internal diameter and 26 cm height, was produced. Denoting the wavelength corresponding to the natural electrical oscillation in air $\lambda$ (where $\lambda = 3 \times 10^{10}\, T$, and $T$ is the self-resonance period), it was found that, if the coil was in air, $\lambda/2 = 276$ cm [$f_0 = 54.3$ MHz]. But when the coil was lowered into an 11 cm wide glass container filled with petroleum, it was found that $\lambda/2 = 360$ cm [$f_0 = 41.6$ MHz]. The ratio of the wave lengths $360/276 = 1.31$ is somewhat smaller than the square root of the dielectric constant of petroleum ($\sqrt{\varepsilon} = 1.41$) because part of the coil (2 cm long) was still sticking out of the petroleum.

- [p304] -

If the coil was removed from the petroleum and pushed onto a glass tube of 15 mm outer diameter and 1.2 mm wall thickness (without changing the pitch or length of the coil), it was found that $\lambda/2 = 290$ cm [$f_0 = 51.7$ MHz]. The small increase of $\lambda/2$ from 276 cm to 290 cm is caused by the small number of the electric field lines from the coil that run in the glass wall parallel to the coil axis. When petroleum was poured into the interior of the glass tube, then the self-resonance wavelength λ of the coil was not significantly increased (because the coil is very long and the glass wall is rather thick in comparison to coil diameter); but when distilled water was poured into the glass tube, λ/2 increased to 354 cm [$f_0$ = 42.3 MHz]. — Now the water was poured out again, and the empty glass tube with the wound coil was placed in the middle of a thin 23 cm high, 47 mm wide cylinder of 3 mm thick glass. The half-wavelength was then λ/2 = 320 cm [$f_0$ = 46.8 MHz]. The increase from 290 cm to 320 cm is caused by the electric field lines on the outside of the coil, which partly run in the wall of the outer glass cylinder. Petroleum was then poured into this, and the half-wavelength increased again to λ/2 = 340 cm [$f_0$ = 44.1 MHz]. However, if a larger outer glass cylinder was used (11 cm diameter), filling it with petroleum gave λ/2 = 364 cm [$f_0$ = 41.2 MHz].

---

15 [302-1] **'Elictrischen Schwingungen um einem stabförmigen Leiter, behandelt nach der Maxwell'schen Theorie'** [Electrical oscillations around a rod-shaped conductor, treated by Maxwell's theory], M. Abraham, Ann. Phys. 302(11) 1898 (Wied. Ann. 66). P435-472, see p442.

As another example, a coil which was short compared to its diameter was chosen. 10 turns of 1 mm thick copper wire, which was 2 mm thick including its insulation, were wound onto an ebonite cylinder of 5.9 cm diameter and 2.7 cm height[16]. The individual turns were placed close together, so that the overall height of the coil was 2 cm. The length of the copper wire was 192 cm. A coil of exactly the same length was wound on a good dry oak cylinder of the same dimensions. Both cores were then bored out, so that the coils were compared on hollow cylinders, of l.5 mm wall thickness for ebonite, and 3.5 mm thickness for wood.
- [p305] -
The following self-resonance wavelengths were obtained:

|  | λ/2 / cm | $f_0$ / MHz |
|---|---|---|
| Ebonite hollow cylinder | 365 | 41.1 |
| Wood hollow cylinder | 386 | 38.8 |
| Ebonite solid cylinder | 406 | 36.9 |
| Wood solid cylinder | 440 | 34.1 |

It follows that wood has a larger dielectric constant than ebonite [hard rubber]. Now this is indeed the case, as was established directly by cutting thin 0.5 mm plates, made from the same piece of wood, and comparing them with ebonite plates between the 3 mm diameter[17] holding plates of capacitor[18], which according to a previously described method[19] is operated at the resonance line of a small Blondlot's exciter[20], which generated electrical oscillations of 73 cm wavelength measured in air [411 MHz].

The capacitor showed the greatest capacitance (and significant electrical absorption) for the wood fibres cut perpendicular to its plates, lower capacitance (and no electrical absorption) for fibres of wood cut parallel to its plates, and the smallest capacitance for the ebonite. The dielectric constant of ebonite is only slightly smaller than that of the wood plates cut with fibres parallel, but on the other hand it is substantially smaller than the dielectric constant of the wooden plate cut with fibres perpendicular. The Dielectric constant of the wood is thus greatest with fibres parallel to the capacitor plates, but also still greater than the Dielectric constant of Ebonite when the fibres are perpendicular. This is in agreement with the measurements on wood made by Righi[21] and Mack [22] using electric birefringence [double refraction].

---

16 [304-1] The ends of the coil were held in place by small indentations in the ebonite cylinder.
17 This is an extremely small capacitor. The diameter was possibly 3 cm rather than 3 mm.
18 [305-1] The plates had the following thicknesses:
    Wood, perpendicular to the fibres  0.428 mm
    Wood, parallel to the fibres  0.442
    Wood, parallel to the fibres  0.475
    Ebonite disk  0.465
   They fitted tightly between the capacitor plates
19 [305-2] **Eine methode zur messung der Dielectric. const...**, P. Drude, Annalen der Physik, 297(7) 1897 (Wied. Ann. 61). p466-510.
20 See also Blondlot's original paper '**Sur un nouveau procédé pour transmettre des ondulations électriques le long de fils métalliques, et sur une nouvelle disposition du récepteur**' [On a new method for transmitting electrical waves along metal wires, and a new receiver arrangement]; R Blondlot, Comptes Rendus de l'Académie des Sciences, vol. 114, 1892, p. 283 - 286.

- [p306] -

The latter has observed the two electrical refractive indices, in fir[23] particularly:

$n_1 = 1.75$ , $n_1^2 = 3.06$     perpendicular to the fibres
$n_2 = 2.15$ , $n_1^2 = 4.62$     parallel to the fibres

Specifically[24] I have not measured the dielectric constant of wood; because it would be necessary to immerse it in a liquid of the same dielectric constant, and it would not be possible to assess safely the change to the dielectric constant that the wood undergoes through the capillary action of the liquid. On the other hand I have determined the dielectric constant of ebonite, using this type of null method[25] (by immersion in benzene - acetone mixtures), $\varepsilon = 2.79$, and that was exactly the same value for two ebonite pieces of different origin, which were used in the coil cores, both in the direction parallel to the axis of the ebonite cylinder rather than in the direction perpendicular to the axis.

The electrical absorption of the wood in directions parallel to the fibre was noticeable in the coil[26]: with the coil on the solid wooden cylinder the exciter loop had to be closer (17 cm), than with the coil on the solid ebonite cylinder (the distance was 21 cm from the exciter loop) to obtain equally distinct resonance indication in the vacuum tube. Even the inductive excitation of coil on the thin hollow wooden cylinder was noticeably weaker than with the coil on the ebonite cylinder. ***For the construction of Tesla transformers it is therefore best to avoid wood cores, and preferably to use no cores[27] or cores made from ebonite, or possibly also glass rods (or tubes)[28].***

- [p307] -

When a good conductor is placed in the interior of the coil, the intensity of the excitation is considerably reduced and also the self-resonance wavelength of the coil is shorter. So, in the coil on the hollow wooden cylinder, λ/2 of 386 cm [$f_0 = 38.8$ MHz] decreased to λ/2 = 344 cm [$f_0 = 43.6$ MHz], as a 3 cm high 0.5 mm thick hollow brass cylinder of 52 mm outer diameter was inserted into the coil interior, and at the same time the distance between the coil and the exciter loop had to be reduced from 17 cm to 1 cm to restore clear indication from the vacuum tube. This brass cylinder was also introduced into the hollow ebonite cylinder wound with thinner insulated wire (226.5 cm wire length), resulting in the decrease of $\lambda/2$ from 567 cm [$f_0$ = 26.4 MHz] (without brass cylinder ) to $\lambda/2$ = 415cm [$f_0$ = 36.1 MHz] (with brass cylinder). Both results, both the weakening of excitation and the reduction of the natural period, can be explained by the induced current in the brass cylinder (tertiary) flowing in the opposite direction to the coil current and hence reducing the self-inductance of the coil (as does the secondary current of a transformer).

---

21 [305-3] **Doppelbrechung der electrischen Strahlen [The birefringence of electrical rays]**, A Righi, Ann. Phys., Vol. 291(6) 1895 (Wied. Ann. 55), p389-390.
    Also published as: A. Righi, Mem. R. Acc. della Sc. Bologna 4, 1894. p487.
22 [305-4] **Doppelbrechung der electrischer Strahlen,** K. Mack, Ann. Phys., 290(-) 1895 (Wied. Ann. 54), p342- .
    [See also review '**Double refraction in wood**', W Hallock, Science, Aug. 23, 1895, p239-240.]
23 [306-1] **Doppelbrechung der electrischer Strahlen,** K. Mack, Ann. Phys., Vol 292(12) 1895 (Wied. Ann. 56), p717-732. See page 729.
24 [306-2] The two dielectric constants of ash are crudely judged to have the values specified by Mack for fir.
25 [306-3] See: '**Methode zur bestimmung der Dielectricitätsconstanten fester Körper** [Method for determining the dielectric const. of a solid body]', H. Starke, Ann. Phys. 296(4) 1897 (Wied. Ann. 60), p629-641 ;
    also: '**Experimentel-untersuchung über electrische dispersion einiger Organiscuer Säuren, Ester, und von zehn Glassorten**' [Experimental study of electrical dispersion of some organic acids, esters, and 10 types of glass.], K. F. Löwe, Ann. Phys. 302 (11) 1898 (Wied. Ann. 66), p390-410, 582-596. 1898. See p402.
26 [306-4] For very long thin coils, less; but in shorter ones, more.
27 [306-5] The coil can be supported using some thin ebonite rods, or even thin metal rods.
28 [306-6] Cardboard tubes also absorb to some extent.

When changing the nature of the coil core, the same changes in $\lambda$ occur in these short coils as in the long coil discussed earlier on p303; except that the effects are even clearer because the coil is short and wide, so there are more electric field lines inside the coil (see above p303). If, for example, the coil on the wooden hollow cylinder ($\lambda/2$ originally 386 cm [$f_0$ = 38.8 MHz] ) was pushed onto a 6 cm tall glass beaker of 1 mm wall thickness (with the wooden cylinder still present), so $\lambda/2$ increased to 397 cm [$f_0$ = 37.8 MHz]. When petroleum was poured into the beaker, so $\lambda/2$ increased further to 412 cm [$f_0$ = 36.4 MHz] (as on p304 for the thin coil, the introduction of petroleum into the glass tube had negligible effect). When water was poured into the beaker instead of petroleum, $\lambda/2$ increased still further to 511 cm [$f_0$ = 29.3 MHz].

In the coil on the hollow ebonite cylinder, a wooden core ( hornbeam) that fitted with 1 mm play was inserted; $\lambda/2$ then increased from 365 cm [$f_0$ = 41.1 MHz] to 411 cm [36.5 MHz], i.e., the coil took an intermediate position between the solid ebonite cylinder and the solid wooden cylinder.

- [p308] -

Reducing the length of the coil on the hollow ebonite cylinder, either by winding fewer turns, or the same number of turns of thinner wire, reduces the effect of the inserted wooden core; because for a short coil, the electric field lines run more at the coil surface , i.e., in the ebonite cylinder, as the coil ends get closer together. The following table gives information:

$h$ is the height of the coil (i.e., the distance between the centres of the end turns, see Fig. 5.),
$2r$ is the average diameter (which is found using $2\pi r \cdot n = \ell$, where $n$ is the number of turns and $\ell$ is the length of the coil wire).
$\lambda_1$ is the self-resonance wavelength of the coil on the hollow ebonite cylinder,
$\lambda_2$ is the self-resonance wavelength after insertion of the wooden core.

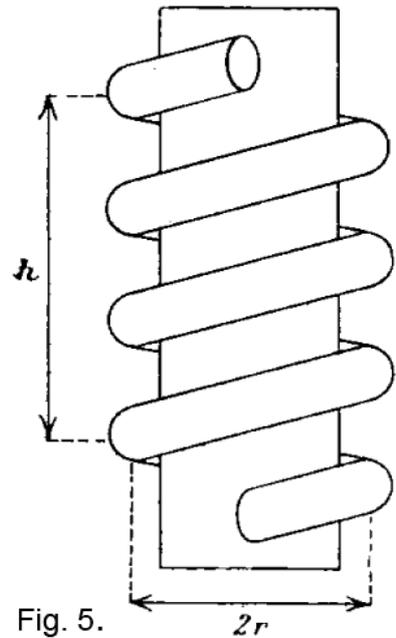

Fig. 5.

**Effect of a wood core in a hollow ebonite cylinder.**

| $h$ / cm | $h/2r$ | $\lambda_1$ | $\lambda_2$ | $\lambda_2 / \lambda_1$ |
|---|---|---|---|---|
| 2.0 | 0.32 | 365 | 411 | 1.13 |
| 1.2 | 0.20 | 567 | 627 | 1.10 |
| 1.0 | 0.16 | 508 | 549 | 1.08 |
| 0.66 | 0.11 | 402 | 417 | 1.04 |
| 0.55 | 0.09 | 360 | 379 | 1.05 |

## 4. Effect of wire insulation on the self-resonance period of the coil

*Thin silk insulation exerts no influence on the natural period of the coil, whereas thicker insulation increases the natural period by 1 - 4 %, specifically, the more so the shorter the coil relative to its diameter.* For example, a coil of height $h$ = 14.9 cm, with $n$ = 48 turns of 1 mm thick bare wire of length $\ell$ = 461 cm, with an exactly constant pitch[29] of 3.17mm, was wound on a wooden core of 2.96 cm diameter; the self-resonance half-wavelength was then $\lambda/2$ = 347 cm.

- [p309] -

Now this wire was unwound and replaced by a 1 mm thick wire with waxed cotton double-insulation, of 2.1 mm thickness overall; the length was again $\ell$ = 478 cm, and $\lambda/2$ = 368 cm. Hence:

|                | $\lambda/2 : \ell$ | $h:2r$ |
|----------------|--------------------|--------|
| Plain wire     | 0.753              | 4.87   |
| Insulated wire | 0.770              | 4.71   |

The ratio $h:2r$, the coil height to the coil diameter, is now not the same in the two cases; and since $\lambda/2:\ell$ depends on this relationship, this must be considered in order to assess the effect of the wire insulation alone. Correcting for this (as explained below[30]) gives:

|                | $h/2r$ | $\lambda/2\ell$ | $p/\%$ |
|----------------|--------|-----------------|--------|
| Plain wire     | 4.87   | 0.753           |        |
| Insulated wire | 4.87   | 0.767           | 1.8    |

i.e., the sole effect of the wire insulation causes a 1.8 %. increase in the ratio $\lambda : 2\ell$.

This effect could be verified in another way:

An ebonite cylinder of 2.72 cm diameter was wound, on the lathe, with a 0.4 mm thick copper wire with thin silk insulation and a 0.6 mm thick cotton thread which lay just between the turns of wire. After the self-resonance period was determined, the thread was unwound while the wire turns exactly retained their original positions. This always resulted in a significant reduction in $\lambda/2$. The results were as follows:

- [p310] -

| $n$ | $h/2r$ | $\lambda/2\ell$ | $p/\%$ |
|-----|--------|-----------------|--------|
| 55  | 2.1    | 1.077  with thread<br>1.060  without thread | 1.6 |
| 29  | 1.0    | 1.415  with thread<br>1.395  without thread | 1.4 |
| 13  | 0.2 [31] | 2.42  with thread<br>2.33  without thread | 3.8 |

*Thus cotton insulation, which is about as thick as the wire, increases the self-resonance period in coils which are at least as high as wide by about 1.5 %, and with shorter coils more* (i.e., about 4%). It is assumed that the insulated wire turns touch each other, or at least that their distance is not great. - This result can be easily understood from section 3, since the insulation has a greater dielectric constant than air.

---

29 [308-1] For this purpose, a shallow thread was cut in the wood core on the lathe.
30 Explanation of the correction procedure is missing from the original paper.
31 [310-1] This observation refers to a thick ebonite cylinder of 5.70 cm diameter.

## 5. Coils with uneven pitch

Six turns of 1 mm thick bare copper wire, with a constant pitch of 5 mm, were wound on an oak core of 12.7 cm diameter. It was found that $\lambda/2 = 462$ cm [$f_0 = 32.4$ MHz]. The wire ends were held while the middle turns were compressed to 3 mm pitch and the pitch at the end coils increased; this increased $\lambda/2$ to 554 cm [$f_0 = 27.1$ MHz]. However, when the end coils were compressed to 3 mm pitch, while the pitch of the middle turns increased, $\lambda/2$ changed to 444 cm [$f_0 = 33.8$ MHz]. *At a fixed coil height* h *and wire-length* $\ell$, *a coil with narrowed central turns has a slower oscillation period, and a coil with narrowed end turns a faster oscillation period, compared to a coil of constant pitch*. This result can be easily understood on the basis that the capacitance of the coil depends essentially only on the coil height *h*, it being created by the electric field lines spanning from the one end of the coil to the other, whereas the self-inductance of the coil arises from the current-carrying turns in the middle.

- [p311] -

Thus, if the pitch *g* is decreased while *h* remains constant, the self-inductance of the coil increases, whereas the capacitance remains constant[32]; hence $\lambda$ must increase. In order to obtain definite conditions, the coil should therefore be made with constant pitch, as is, in practice, always the case in coils of insulated wire with the turns pushed close together. The following considerations relate only to coils wound with constant pitch. The constancy of pitch was achieved either by carefully cutting a coarse thread into the wood core (not deep) on the lathe, or (for insulated wire) by pushing the turns close together.

## 6. The number of characteristic parameters a coil of constant pitch

The following parameters are required for a coil of constant pitch surrounded by air:
- *n* number of turns,
- *g* pitch [ganghöhe],
- *h* coil height,
- 2*r* coil diameter,
- $\ell$ wire length,
- $\delta$ wire thickness, also thickness and type of wire insulation,
- $\varepsilon$ dielectric constant of the core, also its thickness if it is a hollow cylinder.

By separate observations, it was found that the location of the coil on a longer core (whether in the centre of the core or at the end) had no effect; and this also applied to the material of the support on which the coil core rested[33], at least if this support was also made of insulating material (wood).

Now to address the question of how the self-resonance period *T* of the coil, or the self-resonance wavelength $\lambda$, is determined by the parameters of the coil. The parameters are not all independent, because there are the following relations:

$$h = (n-1) g \ , \quad \ell = 2r \pi n$$

According to the similarity rule given earlier on p302, $\lambda$ must now grow in proportion to the length of wire $\ell$ if *n* remains constant, whereas *h, r, $\ell$, g* and $\delta$ grow in the same proportion.

---

32 [311-1] Sometimes the capacitance also increases, because the end turns become closer to some extent even if h remains constant.

33 [311-2] $\lambda$ also remained the same when the coil was not placed on a support, but kept away from other objects by being hung up. This [lack of effect of nearby objects] does not apply to coils without cores, see below.

- [p312] -

It must therefore be that[34]:

(A)  $\lambda/2 = \ell \cdot f(n, h/2r, g/\delta, \varepsilon)$,

Where $f$ is a function of $n$, $h/2r$, $g/\delta$, and $\varepsilon$, and is also somewhat dependent on the nature and thickness of the wire insulation.

First, in order to examine the dependence of $n$, the following coils have been tested with solid ebonite cores:

| $2r$ / cm | $h$ / cm | $\delta$ / mm | $g$ / mm | $n$ | $\ell$ / cm | $\lambda/2$ / cm | $f$ | $h/2r$ | $g/\delta$ | |
|---|---|---|---|---|---|---|---|---|---|---|
| 2.77 | 0.57 | 0.4 | 0.52 | 12 | 105 | 278 | 2.65 | 0.206 | 1.29 | } thin insulation |
| 5.75 | 1.11 | 0.4 | 0.51 | 23 | 415 | 1100 | 2.65 | 0.193 | 1.27 | |
| 5.90 | 1.22 | 0.9 | 2.03 | 7 | 130 | 307 | 2.37 | 0.207 | 2.26 | } thick insulation |
| 5.75 | 1.19 | 0.4 | 0.99 | 13 | 235 | 569 | 2.42 | 0.207 | 2.47 | |

As will be explained later, $f = \lambda/2 : \ell$ is by far mainly dependent on $h/2r$. In the second observation $h/2r$ is slightly smaller than the other observations. Reducing[35] this second observation to the common ratio $h/2r = 0.206$, the value $f = 2.64$ is obtained. Further, as is already evident from this table, $f$ is somewhat dependent on $g/\delta$, in that it decreases from 2.65 to 2.40 when $g/\delta$ increases from 1.3 to 2.4. This decrease of $f$ is, as later experiments showed, much greater near $g/\delta = 1$ than for larger values of $g/\delta$, so that, after reducing to the same values of $h/2r$ and the same values of $g/\delta$, the following result is obtained:

| $n$ | $h/2r$ | $g/\delta$ | $f$ |
|---|---|---|---|
| 12 | 0.206 | 1.29 | 2.65 |
| 23 | 0.206 | 1.29 | 2.63 |
| 7 | 0.207 | 2.26 | 2.37 |
| 13 | 0.207 | 2.26 | 2.43 |

- [p313] -

Increasing the number of turns $n$ from 12 to 23 has therefore reduced $f$ by less than 1%; whereas the increase of n from 7 to 13 has increased $f$ by 2.5%. The last observation is however not exactly comparable to the penultimate, because the insulation material was slightly different in both cases[36].

For larger values of $n$, the dependence of $f$ on $n$ is still insignificant, and always remains below 1%; as is demonstrated by the following table, which refers to coils with wooden cores. The same thin insulated wire was used in all three cases.

| $2r$ / cm | $h$ / cm | $\delta$ / mm | $g$ / mm | $n$ | $\ell$ / cm | $\lambda/2$ / cm | $h/2r$ | $\lambda/2$ reduc. | $g/\delta$ |
|---|---|---|---|---|---|---|---|---|---|
| 1.91 | 8.05 | 0.4 | 1.40 | 58.5 | 350 | 302 | 4.21 | 302 | 3.5 |
| 2.30 | 9.51 | 0.4 | 2.30 | 48.5 | 350 | 304 | 4.14 | 303 | 5.8 |
| 2.97 | 11.57 | 0.4 | 3.17 | 37.5 | 350 | 301 | 3.90 | 295 | 8.0 |

---

34 [312-1] There is a factor ½ attached to $\lambda$, because then, in a straight thin wire in air, $f = 1$.
35 Drude uses 'reducirt' not in the sense of 'making smaller' but in the mathematical sense of 'data reduction', i.e., compacting a data set by reducing the number of parameters associated with it.
36 [313-1] In the latter case, the wires were insulated from each other by a cotton thread. If wound the same, then $f$ for the $n = 13$ case was 1.5 % smaller than for $n = 7$.

When $\ell$ is constant, $\lambda/2$ is therefore almost constant, i.e., independent of n.  The increase of $g/\delta$ causes only a slight decrease in reduced half-wavelength ($\lambda/2$ reduc.) for the same $h/2r = 4.21$.

**The self-resonance period of a coil is therefore independent of the number of turns**, and so we have:

(B)    $\lambda/2 = \ell \cdot f(h/2r, g/\delta, \varepsilon)$.

The following series now refers to two different values of constant $g/\delta$ and changing $h/2r \cdot p$ represents the percentage increase of $f = \lambda/2 : \ell$, when $h/2r$ is held constant and $g/\delta$ changes from 2.4 to 1.09.

- [p314] -

**Ebonite core**

| 2r / cm | g/δ = 2.4, thick cotton insulation[37] | | | 2r / cm | g/δ = 1.09, thin silk insulation | | | p / % |
|---|---|---|---|---|---|---|---|---|
| | n | h/2r | f | | n | h/2r | f | |
| 2.0 to 3.0 | 76 | 5.40 | 0.741 | 2.0 to 3.0 | | | | |
| | 60 | 4.11 | 0.788 | | 107 | 4.11 | 0.808 | 2.5 |
| | 53 | 3.63 | 0.826 | | | | | |
| | 44 | 3.01 | 0.888 | | 79 | 3.01 | 0.924 | 4.0 |
| | 37 | 2.53 | 0.966 | | | | | |
| | 30 | 2.10 | 1.061 | | 54 | 2.10 | 1.110 | 4.0 |
| | 55 | 2.10 | 1.067 | | | | | |
| | 23 | 1.61 | 1.190 | | 42 | 1.61 | 1.233 | 3.6 |
| | 29 | 1.05 | 1.405 | | 29 | 1.05 | 1.521 | 7.9 |
| | | | | | 22 | 0.79 | 1.75 | |
| | | | | | 16 | 0.56 | 2.04 | |
| 5.8 to 6.1 | 10 | 0.32 | 2.11 | 5.8 to 6.1 | | | | |
| | 13 | 0.20 | 2.38 | | 12 | 0.20 | 2.80 | 16 |
| | 7 | 0.20 | 2.38 | | | | | |
| | | | | | 11 | 0.18 | 2.88 | |
| | | | | | 10 | 0.16 | 2.99 | |
| | | | | | 7 | 0.11 | 3.28 | |
| | | | | | 6 | 0.092 | 3.47 | |

From this table it is clearly seen how, on the one hand, at constant $g/\delta$ the function $f$ decreases with increasing $h/2r$, and on the other hand, at constant $h/2r$ the function $f$ increases with decreasing $g/\delta$, and the more so the smaller is $h/2r$.

On **Plate 1**, the results are shown graphically.

In short wide coils, the values of $h/2r$ and $g/\delta$ have such a strong influence that it is necessary to wind the coil very precisely if value of $f$ is to be determined to within 1%.  The table may therefore contain errors of 1 - 2% for $h/2r < 0.6$.

---

37 [314-1] The insulation completely fills the space between the turns.

**Plate 1**

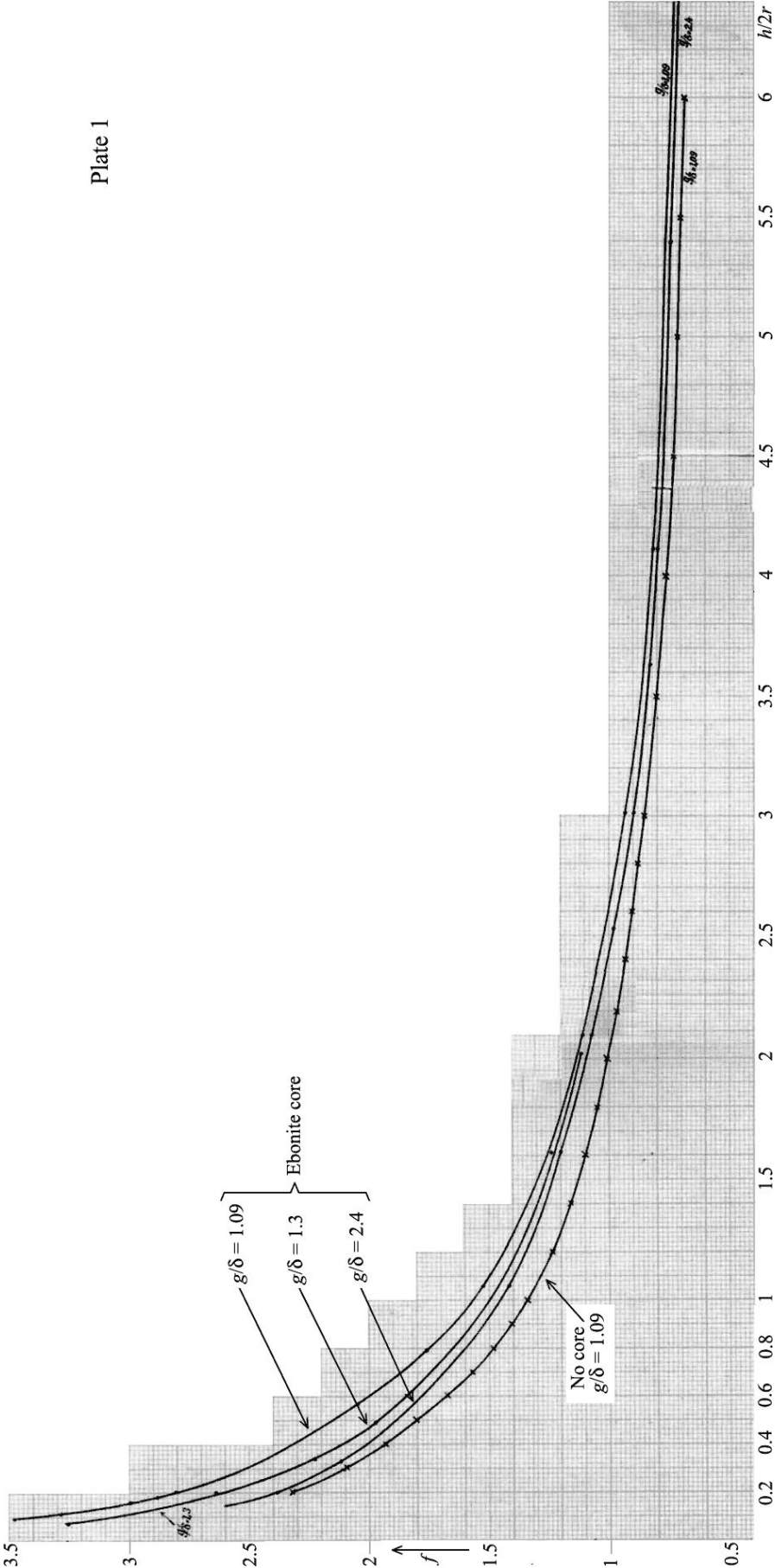

Plate 1

- [p315] -

For $h/2r > 0.6$, the values of $f$ should be accurate to at least 1 %, as is also clear from the smooth course of the curve, and from the fact that when repeating an observation (rewinding the coil)[38] the differences is less than 1%. In the plate there is also a third curve plotted for $g/\delta = 1.27$.

How much $f$ depends on $g/\delta$ at small $h/2r$ is apparent from the following table:

|       | $h/2r = 0.20$ |      |      |      | $h/2r = 2.10$ |      |            |
|-------|------|------|------|------|------|------|------------|
| $g/\delta$ | 1.07 | 1.09 | 1.27 | 2.4  | 1.08 | 1.24 | 2.4 to 2.8 |
| $f$   | 3.00 | 2.80 | 2.64 | 2.38 | 1.12 | 1.10 | 1.06       |

## 7. Coils on wooden cores

As was discussed earlier on p303, the self-resonance wavelength of a coil on a wooden core is greater than that of an otherwise identical coil wound on an ebonite core, and the more so the smaller the value of $h/2r$. In addition, there is an effect due to the the type of wood; good dry (seasoned) cores of ash, beech, hornbeam, and oak were used. The fibres ran parallel to the coil axis.

If we denote $f$ in formula (B) for a wood [holz] core as $f_h$, and for an ebonite core as $f_e$, and define:

$$p = \frac{f_h - f_e}{f_e} \cdot 100$$

as the percentage increase of $f$ in changing from ebonite core to wood core, where $p$ is independent of $g/\delta$. The dependence of $p$ on $h/2r$ and the nature of the wood is represented by the following experimental results:

- [p316] -

|        | $p$ / % |       |          |      |
|--------|------|-------|----------|------|
| $h/2r$ | Ash  | Beech | Hornbeam | Oak  |
| 3.77   | 4.5  | 3.3   | -        | -    |
| 2.00   | 4.5  | 8     | 9        | 9    |
| 1.00   | 7.7  | -     | 11.5     | 12.8 |
| 0.32   | 8.5  | -     | -        | -    |
| 0.20   | 9.7  | 10    | 10.7     | 12.4 |
| 0.10   | -    | 6.3   | -        | -    |
| 0.04   | -    | -     | -        | 12.0 |

The results are not very accurate, because different cores of the same type of wood have slightly different dielectric constants. It is however clear from the table, that ash and beech are about the same, that hornbeam has a greater dielectric constant, and oak the largest. In the latter two types of wood also, $p$ is less strongly affected by decreasing $h/2r$ than in the first two; which would be explained by the assumption that oak and hornbeam are more electrically anisotropic than ash and beech, which means that the dielectric constant in the direction of the fibres is greater than perpendicular to the fibres. In an isotropic material of dielectric constant greater than that of ebonite, for reasons mentioned on p303, $p$ must increase with decreasing $h/2r$. If however the dielectric constant is substantially larger in the direction of the coil axis than in the perpendicular direction, then due to the relatively small number of electric field lines that run parallel to the coil axis on the inside, there must be, at large $h/2r$, a fairly strong increase in the capacitance of the coil, i.e., an increase in $\lambda/2$ will occur.  For smaller $h/2r$, the internal electric field lines of the coil will run partly out of parallel with the coil axis, i.e., in directions having a smaller dielectric constant.  Therefore, the more the dielectric constant of the core is greater in the direction of the axis than in the perpendicular direction, the less will be the increase in $p$ with decrease of $h/2r$.

---

38 [315-1] Some of the values in the table are averages of two separate observations. The observation of the wavelength is accurate to 0.25%. In cases where $f$ has not been determined as precisely, the only reasons are that the coils are not wound sufficiently accurately, and the insulating material of the wire has an effect.

By graphical adjustment, the following values of p have been taken from the table provided, and these are the basis of later calculations.

- [p317] -

$$p = \frac{f_h - f_e}{f_e} \cdot 100 \quad \text{for wood species.}$$

| $h/2r$ | Ash & Beech | Hornbeam | Oak |
|---|---|---|---|
| 6 | 3 | 5 | 6 |
| 5 | 3.5 | 5.5 | 6.5 |
| 4 | 4 | 6.5 | 7.5 |
| 3 | 5 | 7 | 8 |
| 2 | 6 | 8.5 | 9.5 |
| 1.5 | 7 | 9.5 | 11 |
| 1 | 8 | 10.5 | 12 |
| 0.6 | 8.5 | 11 | 12.5 |
| 0.4 | 9 | 11 | 12.5 |
| 0.2 | 9 | 11 | 12.5 |
| 0.1 | 9 | 11 | 12.5 |
| 0.05 | 9 | 10.5 | 12 |

## 8. Coils on hollow cores (tubes)

With coils on hollow cores, apart from the conditions $h/2r$ and $g/\delta$, there is also the ratio $w{:}r$, i.e., wall thickness to radius of the core.  Again, $p$ will be used to represent the percentage increase in $f$ in the transition from ebonite core to hollow core , i.e.,

$$p = \frac{f_h - f_e}{f_e} \cdot 100$$

the result p is independent of $g/\delta$, but dependent on $h/2r$ and $w{:}r$.  The following values of $p$ were observed:

**Coil on ebonite tube**, $w/r = 0.05 = 1/20$

| $h/2r$ | 0.32 | 0.2 | 0.16 | 0.11 | 0.09 | 0.067 |
|---|---|---|---|---|---|---|
| $p\,/\,\%$ | -10.0 | -10.3 | -9.7 | -7.9 | -8.4 | -4.6 |

**Coil on glass tube**, $w/r = 0.05 = 1/20$

| $h/2r$ | 5.36 | 2.00 | 0.64 | 0.33 |
|---|---|---|---|---|
| $p\,/\,\%$ | -6.1 | -6.1 | -6.1 | -4.0 |

- [p318] -

**Coils on glass tubes** (beakers).

|       | $w/r = 1/5$ |      | $w/r = 1/9$ | $w/r = 1/50$ |
|-------|------|------|------|------|
| $h/2r$ | 5.45 | 2.0  | 0.31 | 0.045 |
| $p\,/\,\%$ | -3.4 | -0.9 | +5.6 | +7.1 |

**Coil on cardboard tube**, $w/r = 1/12$

| $h/2r$ | 1.8 |
|---|---|
| $p\,/\,\%$ | -4 |

**Coil on ash-wood tube**, $w/r = 0.11 = 1/9$

| $h/2r$ | 0.32 |
|---|---|
| $p\,/\,\%$ | -4.3 |

It follows from this, as has already been said in section 3, that the natural period is reduced when $w/r$ is small. From the first series of observations listed here, coils on ebonite tubes[39], at a certain value of $w/r$, approach in their natural period that of geometrically similar coils on solid ebonite cores as $h/2r$ becomes smaller. This is also consistent with the observation made on p308, which is that the effect of pushing a wooden core into the ebonite tube is smaller the smaller the value of $h/2r$.

For increasing $h/2r$ at constant values of $w/r$, coils on tubes become more like coils without solid core, and indeed this is more the case as $w/r$ becomes smaller, and also as the dielectric constant of the tube material becomes smaller.

In fact, we see this confirmed by the coils on ebonite tubes (and glass), of $w/r = 1/20$. As we will see in the next section, for a coil of $h/2r = 0.3$ without core, the value $p = -17$ %. For the coil on an ebonite tube with this value of $h/2r$, the value of $p = -10\%$, and for the coil on a glass tube $p = -4\%$.

- [p319] -

For $h/2r = 5.36$, for the coil on a glass tube, $p = -6.1$ %. For a coreless coil is $p = -7.5$ %. For ebonite tube with $w/r = 1/20$, at $h/2r = 5.36$, $p$ must therefore lie between -7.5 and -6.1%, at about $p = -7\%$. Assuming that, between $h/2r = 5.4$ and $h/2r = 0.32$ with the ebonite tube, $p$ changes almost[40] linearly from $p = -7\%$ to $p = -10$ %, we obtain the following tables for $p$ :

**Coils on ebonite tubes**, $w/r = 1/20$

| $h/2r$ | 0.04 | 0.05 | 0.06 | 0.075 | 0.09 | 0.105 | 0.13 | 0.16 |
|---|---|---|---|---|---|---|---|---|
| $p\,/\,\%$ | -3 | -4 | -5 | -6 | -7 | -8 | -9 | -10 |

| $h/2r$ | 0.2 | 0.32 | 1.0 | 1.5 | 2.2 | 3.2 | 4.2 | 6 |
|---|---|---|---|---|---|---|---|---|
| $p\,/\,\%$ | -10.5 | -10 | -9.5 | -9 | -8.5 | -8 | -7.5 | -7 |

**Coils on glass tubes**, $w/r = 1/20$

| $h/2r$ | 0.04 | 0.06 | 0.08 | 0.1 | 0.15 | 0.2 | 0.25 | 0.3 | 0.35 | 0.4 | 0.7 - 6.0 |
|---|---|---|---|---|---|---|---|---|---|---|---|
| $p\,/\,\%$ | +10 | +9 | +8 | +7 | +5 | +3 | 0 | -2 | -4 | -5 | -6 |

---

39 [318-1] The observations on the coils on glass tubes are not as good in comparison with each other, due to variation of dielectric constant with glass type.

40 [319-1] The table was obtained by graphical adjustment [i.e., smoothing]. The error will not exceed 0.5%.

## 9. Coils without cores

A method for producing such coils that worked quite well was first to wind them on a solid core, then remove them carefully and compress the turns together with light pressure by binding them with three pieces of twine, so that a good cylindrical shape was restored.

*In coils without solid* (or liquid) *cores the shortest self-resonance periods are to be expected.* They also work well in fact. Due to the absence of any absorption, the response of the coil at resonance is of course flawless, and also a coil of this construction has the smallest possible capacitance; *so secondary coils without a core are best for Tesla transformers.*

- [p320] -

(There is however the question of how to produce the best coil technically, without it being too easily deformable.)

The percentage change *p* of the coefficients *f* in equation (B) p313 at the transition from coil with ebonite core ($f_e$) to geometrically similar coil without core ($f_0$), will be denoted again by

$$p = \frac{f_0 - f_e}{f_e} 100$$

The following results were obtained (g/δ was either 1.09 or 2.4):

**Coils without core.**

| h/2r | 4.31 | 2.70 | 1.68 | 1.08 | 0.193 |
|---|---|---|---|---|---|
| p / % | -8.4 | -9.1 | -12.3 | -14.5 | -17.1 |

That *p* becomes steadily smaller [more negative?] with decreasing *h/2r* is to be expected, because a coil core increases the natural period more as *h/2r* becomes smaller.

Plotting the values of *p* graphically, we obtain the following representation:

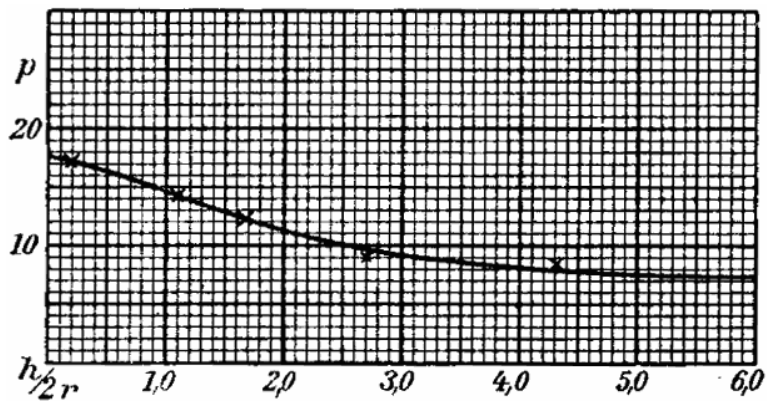

Fig. 6.

The observed values are marked by crosses ×. From this curve, the following table shows the calculation of the coefficients $f_0$ for coreless coils.

$$f_0 = f_e\left(1 - \frac{p}{100}\right)$$

| h/2r | 0.2 | 0.4 | 0.6 | 0.9 | 1.2 | 1.5 | 1.8 | 2.1 | 2.5 | 3.0 | 4.3 | 6.0 |
|---|---|---|---|---|---|---|---|---|---|---|---|---|
| p / % | 17 | 16.5 | 16 | 15 | 14 | 13 | 12 | 11 | 10 | 9 | 8 | 7.5 |

These coils were freely suspended by a cotton thread. ***When they were placed on ebonite, wood or glass, the period was increased as a consequence***, specifically:

| | |
|---|---|
| with $h/2r = 1$ | around 5% resting on ebonite |
| | around 8% resting on wood or glass |
| with $h/2r = 0.2$ | around 4% resting on wood |

If $h/2r$ is very small, the value of $p$ given in the table should only be applied when the wire insulation is not too thick (not larger than the wire thickness), because otherwise the coil will behave as though it is wound on a hollow core, i.e., $p$ will be smaller.

## 10. Tables for calculating the self-wavelength of a coil

The tables set out here for convenient use were obtained by graphical interpolation from the observations of coils on solid ebonite cylinders (see p314), because these can be wound exactly and the material of the coil core is well defined. According to sections 7, 8 and 9, after the values of $f$ were calculated for wood cores and hollow cores and vice versa, the observations of wood and hollow cores were used to supplement the observations of ebonite cores with very small $h/2r$. For wooden cores, observations were also made with large values of $g/\delta$. It is necessary to distinguish three cases:

  a) The turns have no air space between. Turns are in a groove in the ebonite, or are insulated turns pushed together .

  b) The turns have intermediate air space; bare wire in a shallow groove on the core.

  c) Turns of wire in a groove in the wood (without air space between)[41].

  For $g/\delta < 1.3$ only case a) is considered. This is the most important in practice.

  In the table, $h$ is the height of the coil, $2r$ is the coil diameter, $g$ is the pitch , $\delta$ is the wire thickness, and $w$ is the wall thickness of the hollow core.

  The self-resonant half-wavelength of the coil is:

  $\lambda/2 = f \cdot \ell$

where $\ell$ is the length of the coil wire.

---

41 [321-1] If the windings lie in wood, then $f$ is about 2% greater than if the turns are pushed together with cotton insulation. If the corresponding value of $f$ from the tables is not stated directly, it is easy to see that the values of $f$ in the Columns a) need to be enlarged by about 2%.

| | Values of $f$ for coils | | | | | | | | | | | | | | | |
|---|---|---|---|---|---|---|---|---|---|---|---|---|---|---|---|---|
| | Solid ebonite core | | | | without core | | | with ebonite tube $w/r = 1/20$ | | | | with glass* tube $w/r = 1/20$ | | | |
| $h/2r$ | $g/\delta$ | | | | $g/\delta$ | | | $g/\delta$ | | | | $g/\delta$ | | | |
| | 1,09 | 1,24 | 2,4 a) | 2,4 b) | 1,09 | 1,24 | 2,4 a) | 1,09 | 1,24 | 2,4 a) | 2,4 b) | 1,09 | 1,24 | 2,4 a) | 2,4 b) |
| 6,0 | 0,74 | 0,73 | 0,72 | 0,71 | 0,68₅ | 0,67₅ | 0,66₅ | 0,69 | 0,68 | 0,67 | 0,66 | 0,70 | 0,69 | 0,68 | 0,67 |
| 5,5 | 0,75₅ | 0,74₅ | 0,73₅ | 0,72₅ | 0,70 | 0,69 | 0,68 | 0,70₅ | 0,69₅ | 0,68₅ | 0,67₅ | 0,71 | 0,70 | 0,69 | 0,68 |
| 5,0 | 0,77₅ | 0,76₅ | 0,75₅ | 0,74₅ | 0,71₅ | 0,70₅ | 0,69₅ | 0,72 | 0,71 | 0,70 | 0,69 | 0,73 | 0,72 | 0,71 | 0,70 |
| 4,5 | 0,79₅ | 0,78₅ | 0,77₅ | 0,76₅ | 0,73₅ | 0,72₅ | 0,71₅ | 0,74 | 0,73 | 0,72 | 0,71 | 0,75 | 0,74 | 0,73 | 0,72 |
| 4,0 | 0,82₅ | 0,81₅ | 0,80₅ | 0,79 | 0,76 | 0,75 | 0,74 | 0,76₅ | 0,75₅ | 0,74₅ | 0,73₅ | 0,78 | 0,77 | 0,76 | 0,75 |
| 3,5 | 0,87 | 0,86 | 0,84₅ | 0,83₅ | 0,79₅ | 0,78₅ | 0,77 | 0,80 | 0,79 | 0,78 | 0,77 | 0,82 | 0,81 | 0,80 | 0,79 |
| 3,0 | 0,93 | 0,91 | 0,89 | 0,88 | 0,84₅ | 0,83 | 0,81 | 0,85₅ | 0,83₅ | 0,82 | 0,81 | 0,87 | 0,85 | 0,83 | 0,82 |
| 2,8 | 0,96 | 0,94 | 0,91₅ | 0,91 | 0,87₅ | 0,85₅ | 0,83₅ | 0,88₅ | 0,86₅ | 0,84₅ | 0,83₅ | 0,90 | 0,88 | 0,86 | 0,85 |
| 2,6 | 0,99₅ | 0,97 | 0,94₅ | 0,94 | 0,90 | 0,88 | 0,86 | 0,91₅ | 0,89 | 0,86₅ | 0,85₅ | 0,93 | 0,91 | 0,89 | 0,88 |
| 2,4 | 1,03 | 1,00 | 0,98₅ | 0,97 | 0,92₅ | 0,90 | 0,88₅ | 0,94₅ | 0,92 | 0,90 | 0,89 | 0,97 | 0,94 | 0,92 | 0,91 |
| 2,2 | 1,07₅ | 1,05 | 1,03 | 1,01₅ | 0,96 | 0,93₅ | 0,92 | 0,98 | 0,96 | 0,94 | 0,92₅ | 1,01 | 0,99 | 0,96 | 0,95 |
| 2,0 | 1,12₅ | 1,10₅ | 1,08 | 1,06₅ | 0,99₅ | 0,97₅ | 0,96 | 1,02₅ | 1,00₅ | 0,98₅ | 0,96₅ | 1,06 | 1,04 | 1,02 | 1,00 |
| 1,8 | 1,18 | 1,16₅ | 1,13₅ | 1,12 | 1,04 | 1,02 | 1,00 | 1,07₅ | 1,05₅ | 1,03₅ | 1,01₅ | 1,11 | 1,09 | 1,07 | 1,05 |
| 1,6 | 1,25₅ | 1,23 | 1,19 | 1,17₅ | 1,09 | 1,06₅ | 1,04 | 1,13₅ | 1,10₅ | 1,08₅ | 1,06₅ | 1,17 | 1,14 | 1,12 | 1,10 |
| 1,4 | 1,32₅ | 1,29₅ | 1,25 | 1,23 | 1,15 | 1,12 | 1,08₅ | 1,21 | 1,17 | 1,14 | 1,12 | 1,25 | 1,21 | 1,18 | 1,16 |
| 1,2 | 1,42₅ | 1,37₅ | 1,33 | 1,31 | 1,22₅ | 1,18₅ | 1,15 | 1,29₅ | 1,25₅ | 1,21₅ | 1,19₅ | 1,34 | 1,29 | 1,25 | 1,23 |
| 1,0 | 1,55₅ | 1,48₅ | 1,43₅ | 1,41₅ | 1,33 | 1,27 | 1,22₅ | 1,41 | 1,33 | 1,28₅ | 1,26₅ | 1,46 | 1,39 | 1,34 | 1,32 |
| 0,9 | 1,64 | 1,56 | 1,50 | 1,48 | 1,39 | 1,32₅ | 1,27₅ | 1,48 | 1,40₅ | 1,35₅ | 1,33₅ | 1,54 | 1,46 | 1,40 | 1,38 |
| 0,8 | 1,74 | 1,65 | 1,59 | 1,56 | 1,47 | 1,39 | 1,34 | 1,57 | 1,48₅ | 1,43₅ | 1,41₅ | 1,64 | 1,55 | 1,49 | 1,47 |
| 0,7 | 1,85₅ | 1,73 | 1,68 | 1,64 | 1,56 | 1,46 | 1,41 | 1,67 | 1,56 | 1,51 | 1,48 | 1,74 | 1,63 | 1,57 | 1,54 |
| 0,6 | 1,99 | 1,84 | 1,77 | 1,73 | 1,67 | 1,54₅ | 1,48₅ | 1,79 | 1,65 | 1,59 | 1,56 | 1,87 | 1,73 | 1,67 | 1,63 |
| 0,5 | 2,13₅ | 1,95₅ | 1,88 | 1,83 | 1,79 | 1,63₅ | 1,58 | 1,93 | 1,76 | 1,69 | 1,65 | 2,03 | 1,85 | 1,78 | 1,74 |
| 0,4 | 2,30₅ | 2,10 | 2,00 | 1,95 | 1,92₅ | 1,75₅ | 1,67₅ | 2,08 | 1,89 | 1,80 | 1,75 | 2,20 | 2,00 | 1,90 | 1,85 |
| 0,35 | 2,39 | 2,19 | 2,07 | 2,01 | 2,00 | 1,83 | 1,73 | 2,15 | 1,97 | 1,87 | 1,81 | 2,29 | 2,10 | 1,99 | 1,95 |
| 0,3 | 2,50 | 2,29 | 2,15 | 2,08 | 2,08 | 1,91 | 1,79 | 2,25 | 2,06 | 1,94 | 1,87 | 2,45 | 2,24 | 2,11 | 2,04 |
| 0,25 | 2,63 | 2,43 | 2,25 | 2,17 | 2,18 | 2,02 | 1,87 | 2,36 | 2,18 | 2,02 | 1,95 | 2,63 | 2,43 | 2,25 | 2,17 |
| 0,2 | 2,80 | 2,61 | 2,38 | 2,29 | 2,32 | 2,16 | 1,97 | 2,51 | 2,34 | 2,13 | 2,05 | 2,88 | 2,69 | 2,45 | 2,36 |
| 0,15 | 3,04 | 2,80 | 2,53 | 2,43 | 2,52 | 2,32 | 2,10 | 2,74 | 2,52 | 2,28 | 2,19 | 3,20 | 2,94 | 2,66 | 2,55 |
| 0,1 | 3,36 | 3,10 | 2,74 | 2,63 | 2,79 | 2,57 | 2,28 | 3,11 | 2,86 | 2,53 | 2,43 | 3,6 | 3,3 | 2,95 | 2,8 |
| 0,09 | 3,47 | 3,19 | 2,74 | 2,63 | 2,88 | 2,65 | 2,27 | 3,23 | 2,97 | 2,55 | 2,45 | 3,7 | 3,4₅ | 2,95 | 2,8 |
| 0,08 | 3,53 | 3,22 | 2,73 | 2,62 | 2,93 | 2,67 | 2,26 | 3,31 | 3,02 | 2,56 | 2,46 | 3,8 | 3,5 | 2,95 | 2,8 |
| 0,07 | 3,57 | 3,23 | 2,71 | 2,60 | 2,96 | 2,68 | 2,25 | 3,38 | 3,05 | 2,56 | 2,45 | 3,9 | 3,5 | 2,95 | 2,8 |
| 0,06 | 3,62 | 3,21 | 2,68 | 2,57 | 3,00 | 2,66 | 2,23 | 3,44 | 3,05 | 2,55 | 2,44 | 3,9₅ | 3,5 | 2,95 | 2,8 |
| 0,05 | 3,60 | 3,18 | 2,64 | 2,53 | 2,98 | 2,64 | 2,19 | 3,46 | 3,05 | 2,53 | 2,43 | 3,9₅ | 3,5 | 2,9 | 2,8 |
| 0,04 | 3,58 | 3,14 | 2,59 | 2,49 | 2,97 | 2,60 | 2,15 | 3,47 | 3,05 | 2,51 | 2,42 | 3,9₅ | 3,4₅ | 2,8₅ | 2,7₅ |

\* Thüringer glass

| $h/2r$ | Values of $f$ for coils on solid cores ||||||||||||||||
| --- | --- | --- | --- | --- | --- | --- | --- | --- | --- | --- | --- | --- | --- | --- | --- |
| | of ash or red beech ||||| of hornbeam ||||| of oak |||||
| | $g/\delta$ ||||| $g/\delta$ ||||| $g/\delta$ |||||
| | 1,09 | 1,24 | 2,4 a) | 2,4 b) | 3,4—5 c) | 1,09 | 1,24 | 2,4 a) | 2,4 b) | 3,4—5 c) | 1,09 | 1,24 | 2,4 a) | 2,4 b) | 3,4—5 c) |
| 6,0 | 0,76 | 0,75 | 0,74 | 0,73 | 0,74 | 0,78 | 0,77 | 0,76 | 0,75 | 0,76 | 0,79 | 0,77 | 0,76 | 0,75 | 0,76 |
| 5,5 | 0,78 | 0,77 | 0,76 | 0,75 | 0,76 | 0,79 | 0,78 | 0,77 | 0,76 | 0,77 | 0,80 | 0,79 | 0,78 | 0,77 | 0,78 |
| 5,0 | 0,80 | 0,79 | 0,78 | 0,77 | 0,78 | 0,82 | 0,81 | 0,80 | 0,79 | 0,80 | 0,83 | 0,82 | 0,81 | 0,80 | 0,81 |
| 4,5 | 0,83 | 0,82 | 0,81 | 0,80 | 0,81 | 0,85 | 0,84 | 0,83 | 0,82 | 0,83 | 0,86 | 0,84 | 0,83 | 0,82 | 0,83 |
| 4,0 | 0,86 | 0,84 | 0,83 | 0,82 | 0,84 | 0,88 | 0,87 | 0,86 | 0,85 | 0,86 | 0,89 | 0,87 | 0,86 | 0,85 | 0,86 |
| 3,5 | 0,91 | 0,89 | 0,88 | 0,87 | 0,89 | 0,93 | 0,91 | 0,90 | 0,89 | 0,90 | 0,94 | 0,91 | 0,90 | 0,89 | 0,90 |
| 3,0 | 0,98 | 0,96 | 0,93 | 0,92 | 0,94 | 0,99 | 0,97 | 0,95 | 0,94 | 0,96 | 1,00 | 0,98 | 0,96 | 0,95 | 0,97 |
| 2,8 | 1,01 | 0,99 | 0,97 | 0,96 | 0,98 | 1,03 | 1,00 | 0,98 | 0,97 | 0,99 | 1,04 | 1,02 | 1,00 | 0,99 | 1,01 |
| 2,6 | 1,05 | 1,02 | 1,00 | 0,99 | 1,01 | 1,07 | 1,04 | 1,02 | 1,00 | 1,03 | 1,08 | 1,06 | 1,04 | 1,02 | 1,05 |
| 2,4 | 1,09 | 1,07 | 1,05 | 1,03 | 1,06 | 1,11 | 1,08 | 1,06 | 1,04 | 1,07 | 1,12 | 1,10 | 1,08 | 1,06 | 1,09 |
| 2,2 | 1,14 | 1,12 | 1,10 | 1,08 | 1,11 | 1,16 | 1,13 | 1,11 | 1,09 | 1,12 | 1,17 | 1,15 | 1,13 | 1,11 | 1,14 |
| 2,0 | 1,20 | 1,18 | 1,15 | 1,13 | 1,16 | 1,22 | 1,19 | 1,17 | 1,15 | 1,18 | 1,23 | 1,21 | 1,19 | 1,17 | 1,20 |
| 1,8 | 1,26 | 1,24 | 1,21 | 1,19 | 1,22 | 1,29 | 1,26 | 1,24 | 1,22 | 1,25 | 1,30 | 1,28 | 1,26 | 1,24 | 1,27 |
| 1,6 | 1,34 | 1,30 | 1,27 | 1,25 | 1,28 | 1,36 | 1,33 | 1,30 | 1,28 | 1,31 | 1,37 | 1,34 | 1,32 | 1,30 | 1,33 |
| 1,4 | 1,42 | 1,38 | 1,35 | 1,33 | 1,36 | 1,45 | 1,41 | 1,37 | 1,35 | 1,38 | 1,47 | 1,43 | 1,40 | 1,37 | 1,41 |
| 1,2 | 1,54 | 1,48 | 1,43 | 1,41 | 1,44 | 1,57 | 1,52 | 1,47 | 1,45 | 1,48 | 1,59 | 1,53 | 0,48 | 1,45 | 1,50 |
| 1,0 | 1,68 | 1,60 | 1,54 | 1,52 | 1,55 | 1,72 | 1,65 | 1,59 | 1,57 | 1,60 | 1,74 | 1,66 | 1,60 | 1,57 | 1,63 |
| 0,9 | 1,77 | 1,69 | 1,62 | 1,59 | 1,63 | 1,81 | 1,73 | 1,66 | 1,63 | 1,67 | 1,83 | 1,74 | 1,68 | 1,64 | 1,71 |
| 0,8 | 1,88 | 1,78 | 1,72 | 1,68 | 1,73 | 1,92 | 1,82 | 1,75 | 1,71 | 1,75 | 1,94 | 1,83 | 1,77 | 1,73 | 1,80 |
| 0,7 | 2,01 | 1,88 | 1,81 | 1,77 | 1,82 | 2,05 | 1,92 | 1,84 | 1,80 | 1,84 | 2,08 | 1,95 | 1,88 | 1,84 | 1,91 |
| 0,6 | 2,16 | 1,99 | 1,92 | 1,88 | 1,93 | 2,21 | 2,04 | 1,96 | 1,91 | 1,96 | 2,24 | 2,07 | 1,99 | 1,95 | 2,01 |
| 0,5 | 2,32 | 2,12 | 2,04 | 1,99 | 2,04 | 2,37 | 2,18 | 2,09 | 2,04 | 2,10 | 2,40 | 2,19 | 2,11 | 2,06 | 2,12 |
| 0,4 | 2,52 | 2,30 | 2,19 | 2,13 | 2,18 | 2,56 | 2,35 | 2,24 | 2,18 | 2,23 | 2,60 | 2,37 | 2,26 | 2,20 | 2,26 |
| 0,35 | 2,60 | 2,39 | 2,26 | 2,19 | 2,25 | 2,65 | 2,43 | 2,31 | 2,24 | 2,30 | 2,69 | 2,46 | 2,33 | 2,26 | 2,33 |
| 0,3 | 2,72 | 2,50 | 2,34 | 2,26 | 2,32 | 2,78 | 2,54 | 2,39 | 2,31 | 2,37 | 2,82 | 2,59 | 2,43 | 2,35 | 2,42 |
| 0,25 | 2,88 | 2,65 | 2,46 | 2,37 | 2,43 | 2,93 | 2,68 | 2,50 | 2,41 | 2,47 | 2,97 | 2,74 | 2,54 | 2,45 | 2,52 |
| 0,2 | 3,05 | 2,82 | 2,59 | 2,49 | 2,55 | 3,12 | 2,86 | 2,64 | 2,54 | 2,60 | 3,16 | 2,95 | 2,69 | 2,59 | 2,66 |
| 0,15 | 3,32 | 3,05 | 2,76 | 2,65 | 2,71 | 3,38 | 3,11 | 2,81 | 2,70 | 2,76 | 3,43 | 3,15 | 2,85 | 2,74 | 2,80 |
| 0,1 | 3,7 | 3,4 | 3,0 | 2,85 | 2,85 | 3,7 | 3,45 | 3,05 | 2,9 | 2,85 | 3,8 | 3,5 | 3,1 | 2,95 | 2,9 |
| 0,09 | 3,8 | 3,5 | 3,0 | 2,85 | 2,8 | 3,85 | 3,55 | 3,05 | 2,95 | 2,8 | 3,9 | 3,6 | 3,1 | 2,95 | 2,85 |
| 0,08 | 3,85 | 3,5 | 3,0 | 2,85 | 2,75 | 3,9 | 3,6 | 3,05 | 2,95 | 2,75 | 3,95 | 3,6 | 3,05 | 2,95 | 2,8 |
| 0,07 | 3,9 | 3,5 | 2,95 | 2,85 | 2,75 | 3,95 | 3,6 | 3,0 | 2,9 | 2,75 | 4,0 | 3,6 | 3,05 | 2,9 | 2,8 |
| 0,06 | 3,95 | 3,5 | 2,9 | 2,8 | 2,75 | 4,0 | 3,55 | 2,95 | 2,85 | 2,75 | 4,05 | 3,6 | 3,0 | 2,9 | 2,8 |
| 0,05 | 3,95 | 3,5 | 2,9 | 2,75 | 2,75 | 4,0 | 3,5 | 2,9 | 2,8 | 2,75 | 4,05 | 3,55 | 2,95 | 2,85 | 2,8 |
| 0,04 | 3,9 | 3,45 | 2,85 | 2,7 | 2,7 | 3,95 | 3,5 | 2,85 | 2,75 | 2,7 | 4,0 | 3,5 | 2,9 | 2,8 | 2,75 |
| 0,03 | 3,85 | 3,4 | 2,75 | 2,6 | 2,6 | 3,9 | 3,4 | 2,75 | 2,65 | 2,6 | 3,95 | 3,45 | 2,8 | 2,7 | 2,65 |

The data in the table on p322 are more reliable than those in the table on p323, in which there may be errors due to varying wood texture. The most reliable data for $f$ are for solid ebonite cores; where for $h/2r > 0.3$, the accuracy is at least 1%, and for $h/2r < 0.3$ it is at least 2%. For wood cores of small $h/2r$ ($< 0.1$) deviation of the data in the table is possibly as much as 5% due to variable wood texture, but in general, including wood cores, the deviations of the data in table are within 2%. For tubes of smaller wall thickness than $w/r = 1/20$, $f$ naturally lies between the values that the table gives for tubes with $w/r = 1/20$ and for coreless coils.

## 11. Approximate theory of self-oscillation of a long thin coil

When the current in the coil is constant, then (for large values of $h/2r$ strictly, and at least approximately with smaller $h/2r$) the self-inductance of the coil is:

(1) $\qquad L = 4\pi q \dfrac{n^2}{h}$,  [cgs]

where $q = \pi r^2$ is the coil cross section. Therefore [since the wire length $\ell \approx \pi\, 2r\, n$]:

$\qquad L = \ell^2 / h$

There is still a factor $(2/\pi)$, which is smaller than 1, to be multiplied-in; because the current intensity in the centre of the coil has its maximum value, while it is zero at the ends. Therefore:

(2) $\qquad L = \dfrac{2\,\ell^2}{\pi h}$

The capacitance of the coil can be evaluated in the following way: Electric charge migrates to the coil ends. Let us imagine the charge $\pm e$ on two squat cylinders (but whose height may include several turns) lying at the ends of the coil; with the separation of the cylinders equal to the coil height $h$, and their radii equal to the coil radius $r$. If we view these short cylinders as infinitely thin circular rings (circles) of radius $r$, the potential is easily computable. In the centre of the circle we construct a line of length $a$ perpendicular to the plane of the loop (Fig. 7) and perpendicular to this a line of length $r'$.

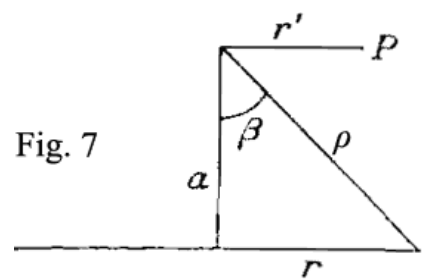

Fig. 7

At the end point $P$ of this line is then the potential generated by the circular line, which can be calculated using spherical harmonics from the formula:

$$V = 2\pi \mathrm{e} \sum_{n=0}^{\infty} \left(\dfrac{r'}{\rho}\right)^n P_0^{(n)} P_\mu^{(n)}$$

where $\mathrm{e}$ is the charge per unit length of the circular line, and

$$\mu = \cos\beta = \dfrac{a}{\rho}, \qquad \rho^2 = a^2 + r^2.$$

Now all spherical harmonics of argument zero of odd order *n* are equal to zero :

$$P_\mu^{(n)}=0, \text{ if n is odd.}$$

Furthermore:

$$P_\mu^{(0)}=1$$

$$P_\mu^{(2)}=\frac{3}{2}\mu^2-\frac{1}{2}$$

$$P_\mu^{(4)}=\frac{5}{2}\cdot\frac{7}{4}\mu^4-\frac{3}{2}\cdot\frac{5}{2}\mu^2+\frac{1}{2}\cdot\frac{3}{4}$$

$$P_\mu^{(6)}=\frac{7}{2}\cdot\frac{9}{4}\cdot\frac{11}{6}\mu^6-\frac{5}{2}\cdot\frac{7}{2}\cdot\frac{9}{4}\mu^4+\frac{3}{2}\cdot\frac{5}{4}\cdot\frac{7}{2}\mu^2-\frac{1}{2}\cdot\frac{3}{4}\cdot\frac{5}{6} \quad \text{etc.}$$

If the point *P* is very close[42] to the circle, then we have:

$$a=0, \quad r'=r=\rho, \quad \mu=0,$$

hence:

$$V=2\pi\text{e}\left\{1+\frac{1}{4}+\frac{9}{64}+\ldots\right\}=2\pi\text{e}.2 \quad ,$$

or , if we introduce the charge *e* of the whole circle:

$$e=2\pi r\cdot\text{e} \quad ,$$

$$V=\frac{2e}{r}$$

- [p326] -

This component of the potential occurs in the coil in addition to the component that results from the -e charged circle at a distance *a* = *h* . Because *r'* = *r* and *a* = *h* , this component is given by:

$$V'=-\frac{e}{r}\sum_{n=0}^{\infty}\frac{1}{(1+h^2/r^2)^n}\ P_0^{(2n)}P_\mu^{(2n)} \quad ,$$

in which

---

42 [325-1] Inside the circle itself, the series for *V* would be divergent, as there is a hypergeometric series ($\alpha = \beta = \frac{1}{2}$ , $\gamma = 1$, x = 1) and this diverges, see Gauss's work on the hypergeometric series, section 15. In reality, of course, we do not have $V = \infty$, because the charge is not on an infinitely thin circular line. The finite size of the charge substituted is therefore reasonable, because *P* only becomes large near the circle. Then the series for *V* has approximately the value $V = 2\pi\text{e}.2$, strictly $V = 2\pi\text{e}.1.9$.

$$\mu^2 = \frac{1}{1+r^2/h^2} \quad .$$

Therefore, the total electric potential at one end of the coil is

$$V_1 = V + V' = \frac{e}{r}\left\{2 - \sum_{n=0}^{\infty} \frac{P_0^{(2n)} P_\mu^{(2n)}}{(1+h^2/r^2)^n}\right\} \quad ,$$

and at the other end of the coil the potential is

$$V_2 = -V_1$$

The potential difference between the coil ends is therefore

$$V_1 - V_2 = 2V_1 = \frac{e}{C} \quad ,$$

Where $C$ is the capacitance of the coil. Therefore

$$C = \frac{r}{2\left\{2 - \sum_{n=0}^{\infty} \dfrac{P_0^{(2n)} P_\mu^{(2n)}}{(1+h^2/r^2)^n}\right\}}$$

The number of factors is now in need of correction because of the assumption that the entire charge of the coil should be concentrated on two circles at the ends. As the charges of the coil are distributed, not on two circular lines, but on several turns of wire, one can think of it as having been replaced by two circular cylinders of finite width, and so the capacitance will be somewhat larger than in the above formula. We can therefore put

(3) $$C = \frac{\alpha r}{2\left\{2 - \sum_{n=0}^{\infty} \dfrac{P_0^{(2n)} P_\mu^{(2n)}}{(1+h^2/r^2)^n}\right\}} \quad ,$$

in which

α > 1.

The numerical factor α will be all the greater than 1 as $h/r$ increases, because the charges of the coil will then be spread over more turns of wire.

- [p327] -

Now, although the numerical value of α is not determined with certainty, it is still reasonable to assume that α is somewhat dependent on $h/r$, so we can write:

(4) $\quad C = r\, \varphi(r/h)$ ,

where φ is a function of the ratio $r/h$.

Now the electrical resonance period $T$ of a system of self-inductance $L$ and capacitance $C_m$ is determined (in electromagnetic units) by the Thomson-Kirchhoff formula:

$$T = 2\pi\sqrt{LC_m} \ ,$$

Therefore the resonant wavelength $\lambda$ is given by the formula

(5) $\qquad \lambda = 2\pi\sqrt{LC} \ ,$

where $C$ is the capacitance in electrostatic units. Using the calculated values of $L$ and $C$ here, we get

(6) $\qquad \lambda = 2\pi\sqrt{\dfrac{2\ell^2}{\pi}\dfrac{r}{h}\varphi(r/h)} \ = \ \ell\,\chi(r/h)$

i.e., **the result is the formula (B) of p313 when the dependency of $\lambda$ on $g/\delta$ is ignored**. In fact this is correct at large values of h/2r [for which] the dependence of $f$ on $g/\delta$ is small.

If the spherical harmonics are expanded only to second order (n = l), which is sufficient for $h/r \geq 3/2$, the result of (3) for the coil capacitance formula is:

(7) $\qquad C = 2\alpha r \dfrac{2 + h^2/r^2 + r^2/h^2}{10 + 4h^2/r^2 + 3r^2/h^2} \ ,$

i.e., according to (2) and (5):

(8) $\qquad \dfrac{\lambda}{2} = 2\ell\sqrt{\alpha\pi\dfrac{r}{h}\cdot\dfrac{2+h^2/r^2+r^2/h^2}{10+4h^2/r^2+3r^2/h^2}}$

From this it follows:

(9) $\qquad f = \dfrac{\ell}{2\lambda} \ = \ 2\sqrt{\alpha\pi\dfrac{r}{h}\cdot\dfrac{2+h^2/r^2+r^2/h^2}{10+4h^2/r^2+3r^2/h^2}}$

This formula is compared below with the empirical results for coreless coils with $g/\delta = 1.09$ [see table on p322].

- [p328] -

The following values for $2\sqrt{\alpha\pi}$ are obtained[43] :

| $h/2r$ | 6 | 5.5 | 5 | 4.5 | 4 | 3.5 | 3 | 2.8 | 2.6 | 2.2 | 2.2 | 2 | 1.8 |
|---|---|---|---|---|---|---|---|---|---|---|---|---|---|
| $2\sqrt{\alpha\pi}$ | 4.76 | 4.64 | 4.52 | 4.42 | 4.32 | 4.22 | 4.17 | 4.16 | 4.13 | 4.10 | 4.07 | 4.04 | 4.01 |

| $h/2r$ | 1.6 | 1.4 | 1.2 | 1.0 | 0.9 | 0.8 | 0.7 | 0.6 | 0.5 | 0.4 | 0.35 | 0.3 |
|---|---|---|---|---|---|---|---|---|---|---|---|---|
| $2\sqrt{\alpha\pi}$ | 3.98 | 3.93 | 3.89 | 3.88 | 3.83 | 3.82 | 3.82 | 3.79 | 3.69 | 3.56 | 3.38 | 3.17 |

Thus we see that, as $h/2r$ increases, α also increases somewhat, as expected. Within the interval

$$2.2 \geq h/2r \geq 1.0$$

formula (9) is fulfilled to within 5 %, and the average value of α would be thus:

$$2\sqrt{\alpha\pi} = 3.97 \, , \quad \alpha = 1.26 \, .$$

The theoretical considerations therefore apply approximately. — The agreement is even better for the coils on solid and hollow cores, as is demonstrated in the following table [on p329], in which the values of $f\cdot\sqrt{h/r}$ for $g/\delta = 1.09$ are indicated for coils on different cores. For sufficiently large $h/r$, this is in agreement with formula (9):

(10) $\quad f\cdot\sqrt{h/r} = \sqrt{\alpha\pi}$

In fact it is particularly evident for the coils on wooden cores that the Product $f\cdot\sqrt{h/r}$ is constant over large intervals of $h/r$, *so that the table can well be used for calculating the value of $f$ at any $h/r$ that is not listed in the tables on p322 and 323* [44].

---

[43] The values in the table have some mistakes and rounding errors. Recalculated values are given in Appendix 1.

[44] [328-1] On the other hand, it becomes very practical to adjust the observation error by adjustment* of the value of $f\sqrt{h/r}$. This was partially done in the preparation of the tables on p322 and 323 .

  * [ graphical adjustment, i.e., smoothing the data. ]

$f \cdot \sqrt{h/r}$ with $g/\delta = 1.09$ for coils wound on:

| $h/2r$ | Ebonite | No Core | Ebonite tube $w/r = 1/20$ | Glass tube $w/r = 1/20$ | Ash or Red beech | Horn-beam | Oak |
|---|---|---|---|---|---|---|---|
| 6,0 | 2,57 | 2,38 | 2,39 | 2,42 | 2,65 | 2,70 | 2,73 |
| 5,5 | 2,51 | 2,32 | 2,33 | 2,36 | 2,60 | 2,64 | 2,67 |
| 5,0 | 2,45 | 2,26 | 2,27 | 2,30 | 2,54 | 2,58 | 2,61 |
| 4,5 | 2,39 | 2,20 | 2,21 | 2,24₅ | 2,48 | 2,53 | 2,56 |
| 4,0 | 2,33₅ | 2,15 | 2,16 | 2,20 | 2,43 | 2,49 | 2,51 |
| 3,5 | 2,30 | 2,10 | 2,12 | 2,16 | 2,41 | 2,46 | 2,48 |
| 3,0 | 2,28 | 2,07₅ | 2,10 | 2,14 | 2,39 | 2,44 | 2,46 |
| 2,8 | 2,27₅ | 2,06₅ | 2,09 | 2,13₅ | 2,39 | 2,44 | 2,46 |
| 2,6 | 2,27 | 2,05 | 2,08 | 2,13 | 2,39 | 2,44 | 2,46 |
| 2,4 | 2,26 | 2,03 | 2,07 | 2,12₅ | 2,39 | 2,44 | 2,46 |
| 2,2 | 2,25₅ | 2,01 | 2,06 | 2,12 | 2,39 | 2,44 | 2,46 |
| 2,0 | 2,25 | 1,99 | 2,05 | 2,11₅ | 2,39 | 2,44 | 2,46 |
| 1,8 | 2,24 | 1,97 | 2,04 | 2,11 | 2,39 | 2,44 | 2,46 |
| 1,6 | 2,23 | 1,95 | 2,03 | 2,10 | 2,39 | 2,44 | 2,46 |
| 1,4 | 2,22 | 1,92₅ | 2,02 | 2,09 | 2,38 | 2,43 | 2,46 |
| 1,2 | 2,21 | 1,90 | 2,01 | 2,08 | 2,38 | 2,43 | 2,46 |
| 1,0 | 2,20 | 1,88 | 2,00 | 2,07 | 2,38 | 2,43 | 2,46 |
| 0,9 | 2,20 | 1,87 | 1,99 | 2,07 | 2,38 | 2,43 | 2,46 |
| 0,8 | 2,20 | 1,86 | 1,98₅ | 2,07 | 2,38 | 2,43 | 2,46 |
| 0,7 | 2,19₅ | 1,85 | 1,98 | 2,07 | 2,38 | 2,43 | 2,46 |
| 0,6 | 2,18 | 1,83 | 1,96 | 2,05 | 2,37 | 2,42 | 2,45 |
| 0,5 | 2,13₅ | 1,79 | 1,93 | 2,03 | 2,32 | 2,37 | 2,40 |
| 0,4 | 2,06 | 1,72 | 1,85₅ | 1,97 | 2,25 | 2,29 | 2,32 |
| 0,35 | 2,00 | 1,67 | 1,80 | 1,92 | 2,18 | 2,22 | 2,25 |
| 0,3 | 1,93₅ | 1,61 | 1,74 | 1,90 | 2,11 | 2,15 | 2,18 |
| 0,25 | 1,86 | 1,54₅ | 1,67 | 1,86 | 2,03 | 2,07 | 2,10 |
| 0,2 | 1,77 | 1,47 | 1,58₅ | 1,82 | 1,93 | 1,97 | 2,00 |
| 0,15 | 1,66₅ | 1,38 | 1,50 | 1,75 | 1,82 | 1,85 | 1,88 |
| 0,1 | 1,50₅ | 1,25 | 1,39 | 1,61 | 1,64 | 1,65 | 1,70 |
| 0,09 | 1,47 | 1,22 | 1,37 | 1,57 | 1,60 | 1,63 | 1,66 |
| 0,08 | 1,41 | 1,17 | 1,32 | 1,52 | 1,55 | 1,58 | 1,60 |
| 0,07 | 1,33₅ | 1,11 | 1,26₅ | 1,46 | 1,46 | 1,48 | 1,50 |
| 0,06 | 1,25₅ | 1,04 | 1,19 | 1,37 | 1,37 | 1,38 | 1,40 |
| 0,05 | 1,13₅ | 0,94 | 1,09₅ | 1,25 | 1,25 | 1,26 | 1,28 |
| 0,04 | 1,01₅ | 0,84 | 0,98 | 1,12 | 1,11 | 1,12 | 1,13 |
| — | — | — | — | — | 0,94 | 0,96 | 0,97 |

- [p328] -

## 12. Coils of few turns and simple loops
As the tables p322 and 323 show, for some small values of $h/2r$ (i.e., $h/2r$ = 0.08 to 0.05, depending on g/δ and core) $f$ has a maximum.

- [p330] -

There it is also understandable that $f$ must decrease again with constantly decreasing $h/2r$, i.e., a constantly diminishing number of turns, because in a simple circular loop ($n$ = 1) $f$ is much smaller than stated in the last lines of the tables above.

In drawing up these last few lines, coils with few turns (down to 3 - 5) were now already being used.

For even smaller numbers of turns, on wooden cores, the following results were obtained:

| n | h/2r | h / cm | 2r / cm | g / mm | δ / mm | ℓ / mm | λ/2 / cm | f | core |
|---|---|---|---|---|---|---|---|---|---|
| 3 | 0.016 | 0.4 | 24.5 | 2.0 | 0.4 | 230 | 622 | 2.70 | Oak |
| 2 | 0.012 | 0.32 | 27.0 | 3.16 | 0.4 | 170 | 375 | 2.20 | ,, |
| 2 | 0.012 | 0.32 | 27.0 | 3.16 | 1.0 | 170 | 410 | 2.41 | ,, |
| 2 | 0.007 | 0.2 | 27.0 | 2.0 | 0.4 | 170 | 409 | 2.40 | ,, |
| 1 | a) | - | 59.6 | - | 0.4 | 187 | 245 | 1.31 | Red beech |
| 1 | b) | - | 58.6 | - | 0.4 | 183 | 257 | 1.40 | Red beech, wire in groove |
| 1 | c) | - | 59.0 | - | 2.0 | 183 | 257 | 1.40 | Red beech, wire in groove |
| 1 | d) | - | 77.0 | - | 2.5 | 243 | 259 | 1.065 | Air (no core) |

The last four rows a) b) c) d) of this table are based on $n$ = 1, i.e., on the **natural wavelength of a simple loop**[45]. The wire circuit was nearly closed, the distance $\Delta$ of the wire ends was changed from 2 cm to 0.5 cm, without affecting $f$. **Neither** (as the table also shows) **does the oscillation period of a simple circuit depend on the wire thickness**[46].

In case a), the wire lay on a 2.5 cm thick 5.5 cm wide wooden ring; in cases b) and c) in a 0.5 cm deep semicircular groove in this wooden ring. In cases b) and c), $f$ appears to be slightly larger than in case a); because the wire, lying in the groove, is more surrounded by wood, which has a dielectric constant larger than that of air.

---

45 [330-1] Since the wavelengths of these simple loops were much smaller than those of the coils, the measuring capacitor $C$ was used without Petroleum filling.

46 [330-2] That is, just as applies for a straight wire, as long as the wire thickness is negligible compared to the wire length. See: **'Die electrischen Schwingungen um einen stabförmigen Leiter, behandelt nach der Maxwell'schen Theorie'** [Electrical oscillations around a rod-shaped conductor treated according to Maxwell's Theory] M. Abraham, Ann. Phys. 302(11) (Wied. Ann. 66). p 435-472, 1898. See p471.

In case d) the wire was supported only by four thin wooden spokes; this case therefore corresponds to being surrounded just by air, with the loop almost closed[47]. Yet for this $f$ was 6.5 % larger than 1, while for a straight thin wire $f = 1$. **The self-resonance half-wavelength of the nearly-closed thin[48] wire loop is 6.5 % larger than its length**.

The increase of the period of a straight wire by bending it into to a circle is quite understandable, since the self-induction will thereby decrease almost imperceptibly, while the capacitance will increase[49].

---

47 [331-1] The distance between the loop and the exciter level was 65 cm, and even then the intensity of the oscillations in the loop was so large that the vacuum tube was observed to glow at a distance of 1 cm from one end of the wire. – Even if, instead of a vacuum tube, sparking between the pointed wire-ends (which were separated by approximately 0.5 mm) was used as the wave indicator; this yielded $\lambda/2 = 259$ cm, the same value as with the vacuum tube as indicator. Therefore, this does not significantly increase the capacitance of the loop (see earlier, p296).

48 [331-2] The experimental conditions were such that the wire thickness was small enough to give the values obtained for $f$ as those applicable to any thin wires; this follows practically from the tests b) and c), where $f$ is independent of δ. After Abraham (see previous citation, same page), $\lambda/2$ for a straight wire of 2.5 mm thickness and 77 cm length is calculated as 0.85% greater than its length $\ell$. When the wire is circularly bent, the correction is the same as for the straight wire; so therefore for a very thin wire loop it should be put that $f = 1.057$ and not 1.065. But notice that the accuracy of the λ comparison in the tests b) and c) was 0.25% ; therefore, it seems that the value $f = 1.065$ for an infinitely thin wire loop is correct.

49 [331-3] If, however, the wire is bent to form two parallel conductors running next to each other, $f = 1$ will be obtained again, provided that the wires are long enough in relation to their distance; because the self-inductance is then reduced in the same ratio as the capacitance has grown. This was verified using a 423 cm long parallel line (wire distance 2.7 cm), which gave $\lambda/2 = 426$ cm. The difference of 3 cm is caused by the proximity (2.5 cm) of a wooden measuring rod. (The data in the tables are not affected by such errors as the proximity of the the wooden measuring rod). - When bending the wire into a circle however, only the ends of the wires, which are charged but carry no current, approach each other, while the current-carrying middle parts are only slightly changed in shape. Therefore the self-inductance remains unchanged, whereas the capacitance increases.

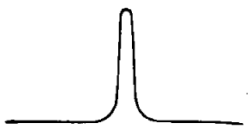

When finally you bend a wire in the form shown in the illustration above, so $f$ must be $< 1$ , i.e., its half-wavelength is shorter than its length, since only the current-carrying middle parts come close, that is, the self-inductance becomes smaller, while the charge-bearing ends do not come near, so that the capacitance does not increase. In fact, a bent wire in this form, in which the ends drawn horizontally in the illustration were each 2 m long, and the vertically-drawn lines were 132 cm long, and the Axial distance was 8 mm (with wire thickness 1 mm ) gave $\lambda/2 = 556$ cm. Since here we have $\ell = 665$ cm , then $f = 556:665 = 0.84$. The measurement method was that first a coil of $\lambda/2 = 556$ cm was placed over the exciter circuit. Then, at one end of the coil, an end of the measuring wire system was connected, and now the 1cm long shorting strap over the parallel section of the wire system was adjusted so that a vacuum tube placed on the free other end of the wire system glowed at maximum brightness.

# 13. A spot-check of the tables using a coil with 23 m self-resonance half-wavelength

As a spot-check of the usefulness of the table; a coil of 0.4 mm thick silk-insulated copper wire, having the following parameters, was wound on a glass cylinder of 1.5 mm thickness and 5 cm diameter:

| $h/2r$ | $h$ / cm | $2r$ / cm | $g$ / mm | $n$ | $\ell$ / cm | $g/\delta$ | $w/r$ |
|---|---|---|---|---|---|---|---|
| 0.89 | 4.55 | 5.1 | 0.49 | 94 | 1500 | 1.23 | 1/17 |

According to the table on p322, when $w/r = 1.20$, $f = 1.47$, i.e., $\lambda/2 = 1:47 \times 1500$ cm = 22.0 m. Such a long wavelength could not be measured easily. The solution however was to replace the petroleum filling of the measuring capacitor $C$ with distilled water. Due to the considerable capacitance increase thereby induced, the Tesla transformer exciter spark could only be produced using a very small spark gap (0.1 mm ?)[50], but with a sufficiently strong excitation of the induction-coil (30 volts) and an increased primary spark gap of the Tesla transformer (3-4 cm) it was possible to produce quite a good exciter spark.

The resonance position of the capacitor showed very well at $d = 9.6$ mm. With Petroleum filling, this $d$ would have a corresponding wavelength $\lambda/2 = 360$ cm (see the curve of Fig. 4 on p301 [moved to p302] ); with water filling therefore, the wavelengths must be multiplied by the ratio 9:1.41, [these being] the square roots of the dielectric constants of water and petroleum[51]. Therefore it follows that

$$\lambda/2 = 360 \cdot \frac{9}{1.41} \text{cm} = 23 \text{m}$$

The agreement of these numbers with the calculated (22 m) has to be called good, especially considering the fact that because of the restriction of the water filling [volume] of the capacitor (see Fig. 2, p295), its capacitance will have increased somewhat less than would be in accordance with the dielectric constant of water (especially as the plate spacing d = 9.6 mm was relatively large), so that $\lambda/2$ must have been slightly smaller than 23 m, and that the calculated value of $\lambda/2 = 22.0$ m, is therefore slightly too small because $w/r = 1/17$, and not 1/20 as the table on p322 implies.

The method used here, that of changing the bath liquid of exciter capacitor $C$, is generally convenient for varying the period over large intervals[52]. With air filling and 5 cm plate distance of the capacitor $C$, $\lambda = 300$ cm = 3 m; with water filling and 1 mm plate distance, $\lambda = 13300$ cm = 133 m.

---

50 [332-1] With direct connection of the exciter wires to the secondary poles of the induction coil only feeble sparks could be produced, even though the striking distance of the induction coil (40 cm) was much greater than the striking distance of the Tesla transformer. The reason apparently is that the voltage between the exciter balls never becomes very high due to the conductivity of the water, because with direct connection of the induction coil they receive only a comparatively slow feed [rise time]. With the interposition of a Tesla transformer, this voltage begins suddenly and is then not so strongly reduced by the conductivity of the water that an exciter spark cannot be produced.

51 [333-1] The Dielectric constant of petroleum, according to footnote [299-2] has been found to be 1.98.

52 [333-2] One such method, by E. Marx (Sachs. Ber. Math.-phys. Kl., Session v. 21 October 1901) has hitherto been portrayed as missing.

In the latter case, it is necessary to use a sufficiently powerful induction coil and Tesla transformer because of the large capacitance of $C$, so that sparking for the oscillatory discharge of $C$ takes place. — If the primary loop is made larger than has been done here (21 cm diameter), the wavelength can of course be further increased.

## 14. Overtones of coils

With sufficiently long coils, overtone resonances can be easily verified using the method given here. With a coil[53] on a solid ebonite cylinder, of $2r = 2.83$ cm and $h = 11.52$ cm, that is, $h/2r = 4.07$, there were three resonance settings of the capacitor $C$; the first (the strongest resonance) gave the wavelength:

$\lambda/2 = 771$ cm (fundamental)

the second:

$\lambda_1/2 = 466$ cm (1$^{st}$ overtone)

the third:

$\lambda_2/2 = 351$ cm (2$^{nd}$ overtone)

That these were overtones could easily be verified by moving the vacuum tube along the coil. There were, on setting the capacitor $C$ to $\lambda/2 = 466$ cm, two null locations on the coil, where the vacuum tube was not lit[54].

For the second overtone, there were three null points, one in the middle of the coil, and two at 1 cm from the coil ends. This distance is less than half the distance between two equivalent nulls (4.76/2 = 2.4 cm), so the potential nodes do not share the coil in equal intervals of $\lambda/4$.

For another coil, of $2r = 2.76$ cm, $h = 5.6$ cm, i.e., $h/2r = 2.02$, n = 114, $\delta = 0.4$, $g = 0.5$, $\ell = 994$ cm, the result was

$\lambda/2 = 1102$ cm (fundamental)
$\lambda_1/2 = 651$ cm (1$^{st}$ overtone)

The overtones are therefore not harmonically related to the fundamental oscillation (apparently because of the strong magnetic coupling between the different parts, each in itself a coil at an overtone). The ratio of the frequencies of the fundamental oscillation to the overtones is somewhat dependent on $h/2r$, since it is

|  | $h/2r = 4.1$ | $h/2r = 2.0$ |
| --- | --- | --- |
| $\lambda : \lambda_1$ | 1.65 | 1.69 |

---

53 [334-1] It had $n = 107$, $\delta = 1$ mm, $g/\delta = 1.09$, $\ell = 953$ cm.

54 [334-2] With intensive magnetic coupling, which is obtained simply by bringing the coil close to the exciter, the coil wire itself lights up. The nodes of the potential variation are then characterised by dark spots on the coil. This is the same phenomenon* that, on a larger scale has recently been described by A. Seibt (Elektrotechn. Zeitschr. 1902. p411. Number 19; Dissert. Rostock 1902).

* [Printing error in the original, should read 'Erscheinung'].

## 15. Increase of the period of coils due to applied capacitance

The parameters for a coil wound on a beech core were:

| h/2r | h / cm | 2r / cm | g / mm | n | δ / mm | ℓ / cm | λ/2 / cm |
|---|---|---|---|---|---|---|---|
| 5 | 15 | 3 | 3.16 | 48 | 1 | 461 | 347 |

at one end of a wire a brass hollow sphere of 7.8 cm in outer diameter was connected, this caused the wavelength to increase from λ/2 = 347 cm to λ/2 = 427 cm. At the same time the potential node moved from the middle of the coil to about 3 cm from the end of the coil to which the sphere was attached, i.e., the potential node was about 4.5 cm from this coil end and about 10.5 cm away from the free coil end. The location of the potential node was again recognized by moving the vacuum tube along the coil while resonance vibrations were generated by it. The vacuum tube is then extinguished at the potential node point.

Likewise, the half-wavelength of 9 cm long coil of 1.8 cm diameter ( $h/2r = 5$ ) increased from λ/2 = 231 cm to λ/2 = 314 cm, when an 18 cm brass plate was connected to the end of the coil.

The change in the period of a coil with capacitance attached at one end can be derived theoretically in the following manner:

The axial direction of the coil is taken as the z coordinate, and the current $i$ at any point in the coil is given by

(11) $\quad i = A . \sin\left\{2\pi \frac{t}{T}\right\} \cos\left\{\frac{\pi z}{2a}\right\}$ .

- [p336] -

The current anti-node [maximum] occurs at $z = 0$ , i.e., the node at $z = a$ is the current node, i.e., the potential anti-node. This is the free end of the coil, while at $z = -a'$ a capacitance $C'$ may be applied. Designating the potential of the coil at an arbitrary point $z$ as $V$; at the coil end $z = -a'$, at which the potential is identical to the potential of the applied capacitance, the following condition must be fulfilled:

(12) $\quad i = -C' \frac{\partial V}{\partial t} \quad$ for $\quad z = -a'$ ,

when the positive current direction is reckoned according to the positive z-axis direction, i.e., directed away from C'.

Designating now, at an arbitrary point $z$, the electric charge present on the length d$z$ of the coil as e. d$z$ , it must thereby result that at this point less current exits than enters. The relation therefore arises:

(13) $\quad -\frac{\partial i}{\partial z} = \frac{\partial e}{\partial t}$ .

On the other hand, for each point $z$ of the coil

(14)  $e = \mathfrak{C} \cdot V$,

where $\mathfrak{C}$ is the capacitance per unit length of the coil (d$z$ = 1) at the point $z$.  From (11), (13) and (14) we obtain

(15)  $\mathfrak{C} \cdot \dfrac{\partial V}{\partial t} = A \sin\left\{2\pi \dfrac{t}{T}\right\} \cdot \dfrac{\pi}{2a} \sin\left\{\dfrac{\pi z}{2a}\right\}$ ,

and therefore it follows that (12) satisfies the condition:

(16)  $\cos\left\{\dfrac{\pi a'}{2a}\right\} = \dfrac{C'}{\mathfrak{C}} \cdot \dfrac{\pi}{2a} \sin\left\{\dfrac{\pi a'}{2a}\right\}$ ,

or

(17)  $\dfrac{\pi}{2a} \cdot \tan\left(\dfrac{\pi a'}{2a}\right) = \dfrac{\mathfrak{C}}{C'}$ .

This equation can first be to used, for an observed $a$ and $a'$, to find the capacitance per unit length $\mathfrak{C}$ of the coil.  In the case above, for example, there was $a = 10.5$, $a' = 4.5$, $C = 3.9$ cm (equal to the radius of the brass ball).  It follows from (17) therefore

$$\mathfrak{C} = 3.9 \cdot \dfrac{\pi}{21} \tan 38.5° = 0.465 \quad ,$$

for the capacitance per unit length of the coil at $z = \pm 4.5$ cm.

Now, according to equation (7), p327, the whole capacitance $C$ of a coil of length $h = 2a'$ is given.  This capacitance $C$ is not spread over the length $a = h/2$ equally however, but rather the elements d$z$ of the coil have more weight the further they are away from the potential node.  The capacitance per unit length $\mathfrak{C}$ of the coil at the point $z = a'$ is therefore obtained by dividing the total capacitance $C$ by a length that is smaller than a', namely the length [given by:]

$$\int_0^{a'} \sin\left(\dfrac{\pi z}{2a'}\right) dz = \dfrac{2}{\pi} a' \quad .$$

Therefore, from equation (7):

(18)  $\mathfrak{C}_{(z=a')} = \dfrac{C}{a'} \cdot \dfrac{\pi}{2} = \alpha \pi \dfrac{r}{a'} \left\{ \dfrac{2 + (2a'/r)^2 + (r/2a')^2}{10 + 4(2a'/r)^2 + 3(r/2a')^2} \right\}$ .

In this case $\alpha = 1.8$ [55], $r = 1.5$, $a' = 4.5$, hence

$\mathfrak{C} = 0.465$ ,

i.e., the calculated value agrees well with that derived from the observations.  Now, although such exact agreement may be somewhat accidental, it shows nevertheless that equation (18) is useful for evaluating the coil capacitance per unit length.

---

55 [337-1] Calculated from the table on p329.

Therefore, the theoretical calculation of the change of λ/2 of a coil by connecting a capacitance on one side is now achieved by first calculating *a*, i.e., the location of the potential node, from (17) and (18) and the overall length *a* + *a*' of the coil. Now λ/2 is very easy to find by noting that the coil must have the same period as a free coil of total length $h = 2a$.

- [p338] -

Therefore, in general, the self-resonance wavelength λ' with attached capacitance C' is given by:

(19) $\quad \lambda'/2 = \ell \cdot \dfrac{2a}{h} \cdot f(2a/2r, g/\delta, \varepsilon)$ ,

and

(20) $\quad \dfrac{\lambda'}{\lambda} = \dfrac{2a}{h} \cdot \dfrac{f(2a/2r, g/\delta, \varepsilon)}{f(h/2r, g/\delta, \varepsilon)}$

where *f* can be taken from the tables on p322 and 323, and *h* is the coil height. Since 2*a* is always greater than *h*, **so the period of a coil with a capacitance C' attached at one end is always increased, but always less than doubled** (i.e., even with C' = ∞, *a* = *h*), because *f*(2*a*/2*r*) < *f*(*h*/2*r*).

With the coil under discussion we had 2*a* = 21 cm, 2*r* = 3 cm and *h* = 15 cm, therefore

$$\dfrac{\lambda'}{\lambda} = \dfrac{21}{15} \cdot \dfrac{0.70}{0.78} = 1.26 \quad ,$$

whereas the observed was λ'/λ =1.23 .

If the attached capacitance C' is very small, so we can put a' = a(1 -ζ) , where ζ is a small number. Then, according to (17)

(21) $\quad C' = \mathfrak{C} \cdot \dfrac{2a}{\pi} \tan\left(\dfrac{\pi}{2}\zeta\right) = \mathfrak{C} \cdot a\zeta = \mathfrak{C} \cdot \dfrac{h\zeta}{2} \quad ,$

where *h* is the coil height.

Furthermore

$$\lambda/2 = f \cdot \ell \quad , \quad \lambda'/2 = f \cdot \ell \cdot 2a/h \quad ,$$

i.e.,

$$\lambda' : \lambda = 2a : h$$

or, because $a + a' = h$ , it follows that

$$2a - a\zeta = h = 2a(1 - \zeta/2) \quad ,$$

$$\frac{\lambda'}{\lambda} = 1 + \frac{\zeta}{2} \quad , \quad \frac{\zeta}{2} = \frac{\lambda'}{\lambda} - 1 \quad .$$

Therefore (21) gives:

(22) $\quad C' = \mathfrak{C} \cdot h \left( \frac{\lambda'}{\lambda} - 1 \right) \quad ,$

or, when the value $\mathfrak{C}$ from (18) is used and $h$ ( $= 2a'$ ) is large compared to $2r$ (the coil diameter):

(23) $\quad C' = \frac{\alpha \pi}{4} \cdot 2r \left( \frac{\lambda'}{\lambda} - 1 \right) \quad ,$

- [p339] -

*Through these equations very small capacitances can easily be determined; e.g., the capacitance increase due to an intensely glowing vacuum tube.*

For example, in a coreless coil of 100 turns of 1 mm thick bare copper wire, of height $h = 30$ cm and diameter $2r = 1.7$ cm, the value $\lambda/2 = 277$ cm was obtained with weak illumination of a vacuum tube at the coil end, against the value $\lambda/2 = 286$ cm with strong illumination (see p297). Therefore the capacitance increase from using the strongly glowing tube, since α is approximately equal to 2, is evaluated as:

$$C' = r\pi \cdot \frac{9}{277} = 0.09 \text{ cm} \quad [0.1 \text{ pF}]$$

(Continued in the next issue.)

(Received, 26$^{th}$ June 1902)

# 4. On the construction of Tesla transformers.
Period of oscillation and self-inductance of the coil.
By P. Drude.
(Continued from p293)

## II. Self-inductance of the coil

For constant currents, Maxwell, Lord Rayleigh , H. Weber and J. Stefan have calculated formulas for the self-inductance of the coil.  For fast current oscillations, the inductance must be smaller.  Instead of calculating this, I am guided by experiment.

## 16. Method of measurement

The method of measurement was that the coils were connected at their ends to the plates of a capacitor *K* of constant capacitance, and then this system was excited inductively using the exciter described on p294.  The distance *d* of the plates of the exciter capacitor was adjusted by micrometer so that a vacuum tube applied to a plate of the capacitor *K* glowed at maximum brightness.  This setting *d* corresponds to the resonance between the exciter and the receiver.  Again, since (as before) the magnetic coupling between the exciter and receiver was chosen to be very weak, so the response was extremely sharp (*d* could be read to 0.02 mm, the measurement accuracy was 0.33%), and the tube only lit up at all if *d* was very close to the resonance position.  — From *d*, using the calibration method described earlier, the wavelength λ of the oscillation was then obtained.

Now, since:

(24)    $\lambda = 2\pi \sqrt{LC}$

if self-inductance is *L*, and *C* is the capacitance (in electrostatic units) of the receiver, the self-inductance *L* follows from λ and *C*.

The Capacitance *C* is calculated[56] using the formula quoted in footnote [299-2].

The capacitor *K* consisted of two circular brass plates[57] of 99.2 mm diameter, between which were three small ebonite plates (squares of 3 mm side length) whose thicknesses were:

    1.013   1.020   1.023   mean: 1.019 mm.

---

56 [591-1] With the fast oscillations, the capacitance of a plate capacitor is slightly less than for static conditions, due to the concentration of the electric field lines towards the edge.  However, this correction is negligible here; see **'Ueber die Periode sehr schneller electrischer Schwingungen'** [The period of very fast electrical oscillations] E. Cohn and F. Heerwagen, Ann. Phys. 279(6) (Wied. Ann. 43). p343-370, see p362. 1891.

57 [591-2] In the centre, two holes of 5 mm diameter were drilled.  The resulting reduction in capacitance is included in the calculations.  One of the capacitor plates was attached by a screw passing through its central hole to an ebonite cylinder of 2 cm in diameter, so that the capacitor could be used in both horizontal and vertical positions.  In the latter case the second capacitor plate was held by three small ebonite brackets.

In a second case, the thicknesses were:

    0.520   0.532   0.538   mean: 0.538 mm

(the thickness of the capacitor plates was 1.5 mm). The capacitance, calculated[58] according to the formula quoted above is:

    $C$ = 63.4 cm  [70.54 pF]

and in the second case is:

    $C$ = 119.5 cm [132.96 pF]

   The experimental determination of $C$ can proceed by connecting simple closed wire loops or wire rectangles to $C$, and then determining $\lambda$. Since it is straightforward to calculate the self-inductance $L$ of circles and rectangles, the capacitance $C$ then follows[59] from $\lambda$.

- [p592] -

This value of $C$ was found[60] to be, in the first case:

    $C$ = 63.1 cm  [70.21 pF]

and in the second case[61]:

    $C$ = 119.1 cm  [132.52 pF]

i.e., smaller than the calculated value C, in the first case by 0.67%, in the second case by 0.33%. Now [regarding] $C$ being somewhat smaller [than calculated], we can be account for this shortfall because the plates were not polished, but had small scratches, which could have 1% impact even when they were just 0.01 mm deep.

   In fact, after polishing the brass plates, this method produced $C$ = 63.5 at 1.019 mm plate spacing, i.e. matching within 0.167% of the theoretical value.

   When using formula (24), it is taken into account that it only strictly applies when the current in the entire closed loop is constant, which is only likely to take place if the wavelength $\lambda$ is very large compared to the wire length $\ell$ of the closed loop. Strictly[62], $\lambda$ is to be calculated from the formula:

(25)    $$\pi \frac{\ell}{\lambda} \tan\left(\frac{\pi \ell}{\lambda}\right) = \frac{\ell^2}{4LC}$$

   If the value of the right-hand side of this equation is small (equate to $a^2$), then $a$ is approximately equal to $\pi \ell / \lambda$. So we can put

---

58 [591-3] The increase in capacitance due to the dielectric constant (2.79) of the ebonite plate is taken into account.
59 [591-4] Formula (24) is not used directly, but with consideration of the correction discussed further below.
60 [In the original, an entry in brackets says ' see below in section 10 ', but that section does not exist.]
61 [592-1] The capacitor was [in parallel with] a 2 mm thick circular loop of length $\ell$ = 54.7 cm. It was found that $\lambda/2$ = 766 cm. The capacitor has not been used for other experiments with such a small plate distance.
62 [592-2] Compare: G Kirchhoff, Ges. Abhandl. p131, 154, 182;  P Drude, Phys. d. Aethers p383, Formula (63).

$$\pi \frac{\ell}{\lambda} = a(1-\delta)$$

and, since: $\tan \varepsilon = \varepsilon + \frac{1}{3}\varepsilon^3$    if $\varepsilon \ll 1$ ,

it follows from (25):

- [p593] -

$$a^2(1-\delta)^2(1+\tfrac{1}{3}a^2) = a^2 = \frac{\ell^2}{4LC}$$

i.e.
$$\delta = \tfrac{1}{6}a^2$$

$$\pi \frac{\ell}{\lambda} = \frac{\ell^2}{2\sqrt{LC}}(1-\tfrac{1}{6}a^2)$$

It is therefore finally

(26)    $$\frac{\lambda}{2} = \pi\sqrt{LC}\left(1 + \frac{\ell^2}{24LC}\right)$$

or

(26′)    $$\frac{\lambda}{2} = \pi\sqrt{LC} + \frac{\pi\ell^2}{24\sqrt{LC}}$$

This equation is used to calculate $\lambda$ from $L$ and $C$. In order to calculate, vice versa, $LC$ from $\lambda$, it follows

(27)    $$\sqrt{LC} = \frac{\lambda}{2\pi}\left(1 - \frac{\pi^2\ell^2}{6\lambda^2}\right)$$

or

(27′)    $$L = \frac{\lambda^2}{4\pi^2 C} - \frac{\ell^2}{12C}$$

## 17. Simple loops

Simple circular loops were made of copper wire of thicknesses $2\rho$ = 1, 2, 3 mm and lengths $\ell$ = 80.8 cm and 54.7 cm.  The capacitor $K$ was used with plate distance d = 1.019 mm.  The wire loops were connected to the capacitor plates by their spring-like action and supported at one point with insulation.  It was found that the connection point of the loops to the capacitor $K$ (whether on the edge or close to the centre) had no effect on $\lambda$.  The following table shows the results.  $r$ is the radius of the circles; from $\lambda$ and from the self-inductance formula

(28) $\quad L = 2\ell\left(\log_e\{8r/\rho\} - 2\right)$

the capacitance $C$ is calculated according to (27) :

Capacitor plates not polished.

| $\ell$ / cm | $2r$ / cm | $2\rho$ / mm | $L$ / cm | $\lambda/2$ / cm | $C$ / cm |
|---|---|---|---|---|---|
| 80.8 | 25.9 | 1 | 908 | 755 | 62.9 |
| 80.8 | 25.9 | 3 | 731 | 677 | 62.7 |
| 54.7 | 17.5 | 1 | 572 | 599 | 63.2 |
| 54.7 | 17.5 | 2 | 498 | 559 | 63.1 |
| 54.7 | 17.5 | 3 | 454 | 536 | 63.5 |

- [p594] -

The mean value is:

$\quad C$ = 63.1 cm [70.21 pF]

The wires were bent almost exactly circular.  This is not critical incidentally, because if a circular wire of length $\ell$ = 54.7 cm was bent into an ellipse with the axis ratio of 3:4, then $\lambda/2$ was reduced by only 0.1%.  On the other hand, it is very closely dependent (within 1 mm) on the wire length $\ell$.  The capacitor circuit was also completed by two shorter wires of $\ell$ = 25 cm in length.  If the capacitance value $C$ = 63.1 cm is used, it follows from (27) that the self-inductance $L$ is ( $L_{calc}$):

| $\ell$ / cm | $2r$ / cm | $2\rho$ / mm | $\lambda/2$ / cm | $L$ / cm obs. | $L$ / cm calc. |
|---|---|---|---|---|---|
| 25.0 | 8.1 | 1 | 374 | 224 | 223 |
| 25.0 | 8.1 | 2 | 342 | 187.5 | 188 |

The observed and the calculated value of *L* are thus identical within 0.5%, i.e., the wires were not so thick in comparison to 2*r* that the uneven distribution of current on the surface of the wire needed to be considered.  Formula (28) assumes that the current flows only on the surface of the wire, with uniform density about the wire axis[63].

When the wire loop is not surrounded on all sides by air, but is wound on an insulating core of greater dielectric constant, then this does not change the self-inductance *L*.

- [p595] -

Nevertheless, the period of the combined capacitor and wire loop can be somewhat larger when the dielectric constant of the core is very large.  It was found, for a 1 mm thick copper wire, which was cemented onto a 15.4 cm wide beaker with sealing wax in two places, that λ/2 = 575 cm.  Distilled water was then poured into the beaker, increasing λ/2 to 585 cm.  On the other hand an alcohol filling gave no noticeable increase in λ/2, and no weakening of the oscillations.  The increase in λ/2 due to the greater dielectric constant of the core is therefore very small, and that is understandable for the following reason: If the length $\ell$ of the wire loop is small compared to the wavelength λ, the current in the wire loop is approximately constant, it decreases only very little at the ends of the wire loop.  This decrease in current is accompanied by an electric charge on the wire surface, i.e. it produces electric field lines.  This increases the capacitance of the whole system, and therefore λ must be somewhat larger than $2\pi\sqrt{LC}$, as formula (26) demonstrates.  If the electric field lines of the closed loop run in a medium of large dielectric constant instead of air, the capacitance of the whole system will increase a little more, i.e., the period continues to increase.  But this increase must be small since it is obviously in the proportion of $\ell^2 : \lambda^2$ as indicated by formula (27).

Therefore, it is safe to use a wood core in the study of coils.  The dielectric constant and dielectric absorption of the wood has no effect as long as $\ell : \lambda$ is not significantly greater than was used in the experiments here ( $\ell : \lambda \leq 0.05$ ); since not even alcohol filling had any influence on λ or the intensity of the oscillations, and yet alcohol has a much greater dielectric constant and larger dielectric absorption, than wood has.

On the other hand, one must be aware of another correction, if the wire loop is wound on a core.  The ends of the wire have to run out for connection to the capacitor plates in two parallel leads that are perpendicular to the surface of the coil core.

- [p596] -

The self-inductance *L'* of the wire ends [see footnote [299-2] ] is

(29)    $L' = 4\ell' \log_e \dfrac{d'}{\rho'}$ ,

where $\ell'$ is the length of each of the two wire ends, *d'* is their axial distance, and 2*ρ'* is the wire thickness. $\ell'$ and $d'$ should of course be made as small as possible, because $L'$ is not completely negligible compared to the self-inductance of the rest of the closed loop.

---

63 [594-1] A rapidly changing current is always distributed so that its self-inductance is a minimum; see. **'Electrische Schwingung in geraden Leitern'** [oscillations in straight conductors], J. Stefan, Ann. Phys. 277(11) (Wied. Ann. 41). p400-420. 1890.  The self-inductance circuit for thicker wires in which *g/r* is not negligible compared to 1, were calculated by Minchin, Electrician 32. p168. 1893 (see. also G. Wiedemann, Lehre von der Elektricität, 2. Aufl., 4. p85. 1898).  It must be somewhat smaller than according to formula (28). — The curvature of the wire causes no appreciable deviation from the formula, cf. **'Ueber die Berechnung und Messung kleiner Selbstpotentiale'** [Calcualation and measurement of small self-potentials],  M Wien, Ann. Phys 289(13) (Wied. Ann. 53) 1894. p928-947, see p935.

In one case, for example, closed loops having a length $\ell = 48$ cm were wrapped around the glass beaker, while two 4 cm long wire leads having a separation of 4 mm led to the capacitor (i.e., $\ell' = 4$ cm, $d' = 0.4$ cm, $g' = 0.05$ cm) giving $\lambda/2 = 579$ cm. The wire leads were then shortened by 2.5 cm (ie, $\ell' = 1.5$ cm), resulting in $\lambda/2 = 565$ cm.

If we call $\zeta$ the percentage increase of $\lambda$, if the self-inductance increases to $L'$, then:

$$(L+L')C = \left(\frac{\lambda'}{2\pi}\right)^2 \left(1 - \frac{\pi^2 \ell^2}{3\lambda^2}\right)$$

$$LC = \left(\frac{\lambda}{2\pi}\right)^2 \left(1 - \frac{\pi^2 \ell^2}{3\lambda^2}\right)$$

$$\lambda' = \lambda(1+\zeta)$$

Thus

$$(30) \quad L'C = 2\zeta \left(\frac{\lambda}{2\pi}\right)^2 \left(1 - \frac{\pi^2 \ell^2}{3\lambda^2}\right)$$

In this case, $\lambda'/2 = 519$ cm and $\lambda/2 = 565$ cm, i.e., $\zeta = 0.025$. Therefore, the self-inductance $L'$ for two 2.5 cm long wires ($d' = 4$ mm, $g' = 0.5$ mm) using the value of $C = 63.1$ cm is:

$L' = 25.5$ cm

while according to formula (29),

$L' = 21$ cm.

The self-inductance $L$ of the closed circuit including two 1.5 cm long leads $\ell'$, is derived from $\lambda/2 = 565$ cm for $L = 510$ cm, while from equations (28) and (29) ($\ell = 48$ cm, $2r = 15.4$ cm, $2\rho = 0.1$ cm, $\ell' = 1.5$ cm, $d' = 0.4$ cm, $2\rho' = 0.1$ cm) $L$ is calculated to be $491 + 13 = 504$ cm.

- [p597] -

A circular loop on a wooden core, with $\ell = 43.8$, $2r = 14$ cm, $2\rho = 0.1$ cm, $\ell' = 1$ cm, $d' = 0.7$ cm, and $2\rho' = 0.1$, showed: $\lambda/2 = 535$ cm. Therefore, it follows from (27) with $C = 63.1$, that $L = 456$ cm, while calculated from (28) and (29) $L = 451$ cm. When a 0.4 mm thick wire was used, then $L = 529$ cm was observed, compared to $L = 532$ cm by calculation. Consequently, and from the table on p593, which indicates the values of $C$, it can be concluded that ***the observation error for $L$ is not more than 1% according to this method***.

## 18. Rectangles

With rectangular closed loops, the self-inductance is also calculated, thereby finding the capacitance of the capacitor from the λ determination. For a rectangle with sides *a* and *b*, the self-inductance (see M. Wien, as cited earlier, p 930) is:

$$(31) \quad L = 4\left\{ a\log_e\left(\frac{2ab}{\rho(a+\sqrt{a^2+b^2})}\right) + b\log_e\left(\frac{2ab}{\rho(b+\sqrt{a^2+b^2})}\right) + 2\left(\sqrt{a^2+b^2} - a - b\right) \right\}$$

If one side *a* is much longer than the other side *b*, and including terms of *b/a* to second order, the result is:

$$(32) \quad L = 4a\left\{ \log_e\left[\frac{b}{\rho}\left(1+\frac{b}{a}\right)\right] - \frac{b}{a}(2-\log_e 2) - \frac{b^2}{4a^2} \right\}$$

or, when the length $\ell = 2(a+b)$ is introduced:

$$L = 2\ell\left\{ \log_e\left(\frac{b}{\rho}\right) - \frac{b}{a}(2-\log_e 2) + \frac{b^2}{a^2}\left(\frac{7}{4} - \log_e 2\right) \right\}$$

i.e.

$$(33) \quad L = 2\ell\left\{ \log_e\left(\frac{b}{\rho}\right) - 1.31\frac{b}{a} + 1.06\left(\frac{b}{a}\right)^2 \right\}$$

2 cm above a millimetre-graduated wooden measuring rod, two taut 1 mm thick copper wires were tensioned in parallel at a separation *b* = 2.65 cm.

- [p598] -

At one end, each wire was affixed by a screw to a plate of the capacitor *K*, through 2 cm long brass connectors[64]; over the other end of the wires a wire shorting-strap *B* could be moved. Depending on the position of *B*, different side lengths *a* were therefore defined for the rectangular loop. The loop was stretched-out 10 cm - 20 cm above the exciter, so that it was excited inductively, and λ was again determined from the resonance. The result were:

| a / cm | λ/2 / cm | L / cm | C / cm |
|---|---|---|---|
| 21.7 | 474 | 362 | 62.6 |
| 27.2 | 530 | 451 | 62.9 |
| 35.7 | 609 | 587 | 63.5 |
| | | | Mean *C* = 63.0 |

This value of *C* is in agreement to within 0.17% with the value obtained using circular loops on

---

64 [598-1] These brass connectors were 4mm thick. Since they did not have the same thickness as the wires, a small correction was applied in the calculation of *L* in accordance with equation (33).

p593. — The capacitor with polished plates resulted in $a = 34.0$, i.e., $L = 565$ cm[65], the half wavelength $\lambda/2 = 598$ cm, giving $C = 63.5$ cm. So this value is within 0.17% of the value of $C$ calculated on p591.

This parallel-wire loop could now be very well used in order to calibrate the exciter for longer wavelengths, where a direct measurement using the method specified on p298 is too inconvenient. One can either proceed in such a way that the plates of the capacitor $C$ are set in the excitation circuit to certain separation $d$, and the position of the shorting-strap $B$ in the parallel line is adjusted to resonance, i.e., the rectangle lengths $a$ are determined, or alternatively, that $B$ is placed at a certain position $a$, and the resonance values are determined from $d$.

- [p599] -

Both methods lead to the same precision. The latter method was chosen, because it was a little more convenient. The calibration results[66] are given in the following table. The value $C = 63.1$ cm is used for capacitor $K$.

| a / cm | d / mm | $\lambda/2$ / cm | $\lambda/2\sqrt{d}$ | $\lambda/2\sqrt{d}$ Smoothed |
|---|---|---|---|---|
| 21.7 | 5.18 | 473 | 1076 | 1076 |
| 27.2 | 4.02 | 529 | 1060 | 1057 |
| 35.7 | 2.97 | 605 | 1042 | 1042 |
| 42.7 | 2.45 | 662 | 1037 | 1037 |
| 53.2 | 1.91 | 738 | 1021 | 1031 |
| 70.7 | 1.44₅ | 851 | 1023 | 1023 |
| 110.7 | 0.93 | 1066 | 1027 | 2019 |

For the reasons previously mentioned in footnote [299-2], $\lambda/2\sqrt{d}$ must decrease somewhat as $d$ decreases, as the table shows. The easiest and most accurate way to derive the half-wavelength $\lambda/2$ from the observed value of $d$ is first to smooth the observed values of $\lambda/2\sqrt{d}$ (column 4 of the table) graphically (using a curve) (→ column 5), thus for any $d$, the corresponding value of $\lambda/2\sqrt{d}$, is taken from the fifth column, and the value $\lambda/2\sqrt{d}$ is divided by the square root of the observed $d$. This, method is followed from here on.

---

65 [598-2] In this experiment, the brass connectors were screwed to the capacitor so that the line anywhere consisted of only 1 mm thick copper wire. Equation (33) was therefore applied without correction.

66 [599-1] The capacitor $C$ was now mounted on metal supports $e\ e$ (see p301) to keep $d$ as steady as possible, and one arm h was grounded. Therefore, the capacitance of the capacitor was slightly larger than it was when using ebonite supports for $e\ e$., and the values of $\lambda/2$ given in the table, for the specified values of $d$, are not in agreement with those from the table on p299, but are larger here by about 15 cm.

## 19. Coils

The wires were fixed to wooden cores (see p595); the wires were fastened at their ends either with wax, sealing wax, or by two 5 mm long, 0.7 mm thick iron-wire pins.

- [p600] -

This showed no detectable influence on the self-inductance, as demonstrated by control experiments that were carried out.

In the following;
$n$ denotes the number of turns,
$2r$ the diameter of the coil,
$h$ is the height of the coil ( $h = (n-1)g$ ),
$g$ is the pitch of the coil,
$\delta$ is the thickness of the coil wire,
$\ell$ is the length of the coil wire in contact with the core,
$\ell'$ the length of the wire-ends that project from the core and lead to the capacitor $K$ (see p596),
$d'$ is their axial distance,
$L_1$ is the total self inductance of the closed loop calculated, using equation (27), from $\lambda$ and the observed value $C = 63.1$ cm bez.[67] $C = 63.5$ cm (see p 598),
$L = L_1 - L'$ is the self-inductance of the actual coil $\ell$, where $L'$ is calculated according to (29).

The observed values of $L/2\ell$ are compared with those calculated by Stefan[68] for slowly changing current values:

$$(34) \quad \frac{L}{2\ell} = n\left\{\left(1 + \frac{h^2 + \frac{1}{8}\delta^2}{32\,r^2}\right)\log_e\left(\frac{8r}{\sqrt{h^2+\delta^2}}\right) - y_1 + \frac{h^2}{16\,r^2}\cdot y_2\right\} + \log_e\left(\frac{g}{\delta}\right)$$

Here, $y_1$ and $y_2$ depend on $\delta/h$ and take their values from a table calculated by J. Stefan, which is reproduced further below. The last term in (34), namely the term $\log_e(g/\delta)$ is not given by Stefan, because his formula for coils of thin insulated wire assumes $g/\delta = 1$ approximately. In the primary winding of Tesla transformers $g/\delta$ must be significantly greater than 1 (eg., thick insulated wire in which $g$ is at least equal to $\delta$ + double insulation thickness), otherwise sparks jump between the turns. Hence, the last term is from Maxwell[69].

---

67 Translation of the abbreviation 'bez.' is not clear here. Drude gets the capacitance of his 1.019 mm spaced capacitor as $C = 63.1$ cm with un-polished plates, and 63.5 cm with polished plates (see p592). It may be that he repeated his experiments after polishing the capacitor, in which case 'bez.' could mean 'respectively' (beziehungsweise), but he only reports his results (table p601) with $C = 63.5$ cm, making the first of the two capacitance values redundant.

68 [600-1] **'Berechnung der Inductionscoëfficienten von Drahtrollen'**, [calc. of induction coeffs. of wire rolls]. J. Stefan, Ann. Phys 258(5) (Wied. Ann. 22). p107-117. 1884. It is assumed here that the wire thickness $\delta$ is small compared to the coil diameter $2r$, which was the case for the coils investigated and will usually be the case as well in practice.

69 [600-2] Cl. Maxwell, Elektricität und Magnetismus 2, deutsch von Weinstein, 2[nd] ed. 2. p407. 1883. See also G. Wiedemann, Lehre von der Elektricität 2, 4[th] ed. p86. Section 119.
[Note that the section and page numbers of the German translation of Maxwell's treatise do not correspond to the original English version.]

| n | ℓ / cm | ℓ' / cm | d' / mm | λ/2 / cm | $L_1$ / cm | L' / cm | L / cm | L:2ℓ obs. | L:2ℓ calc. | L:2ℓ calc - obs |
|---|---|---|---|---|---|---|---|---|---|---|
| | | | | | | | 2r = 2.92 cm, δ = 0.4 mm, g = 2 mm, g/δ = 5, 2r/δ=73, C = 63.5 cm. | | | |
| 2 | 18.3 | 1 | 4 | 397 | 250 | 12 | 238 | 6.50 | 8.38 | 1.88 |
| 3 | 27.5 | 1 | 5 | 539 | 464 | 13 | 451 | 8.20 | 9.96 | 1.76 |
| 4 | 36.7 | 1 | 6 | 668 | 712 | 14 | 698 | 9.51 | 11.29 | 1.78 |
| 5 | 45.8 | 1 | 7 | 794 | 1006 | 14 | 992 | 10.80 | 12.42 | 1.62 |
| 6 | 55.1 | 1.5 | 8 | 901 | 1295 | 22 | 1273 | 11.55 | 13.36 | 1.81 |
| 7 | 64.6 | 1.5 | 9 | 1020 | 1659 | 23 | 1636 | 12.65 | 14.18 | 1.53 |
| 8 | 73.9 | 1.5 | 10 | 1119 | 1997 | 24 | 1973 | 13.36 | 14.88 | 1.52 |
| 9 | 83.1 | 1.5 | 11 | 1211 | 2337 | 24 | 2313 | 13.92 | 15.55 | 1.63 |

The difference:

$$\Delta = L/2\ell \text{ (calc.)} - L/2\ell \text{ (obs.)}$$

therefore decreases slightly with increasing number of turns *n*. This has been observed in all cases (i.e. for other values of g/δ and 2r/δ).

From the numerous other observations, only the final results for the difference Δ and (approximate) observed values of *L*/2ℓ are given. An exclamation mark[70] after the Δ value (eg., Δ = 1.69 (!) for *n* = 2), means that the coil in question has a very uniform pitch because the coil wire was set into a helical groove that was machined by lathe into the coil core. The Δ values without the (!) refer to coils for which this was not the case, i.e., in which the specified value of pitch *g* is not as uniform.

The number above the Δ value is the ratio g/δ, directly below the Δ value is the ratio 2r/δ, and the bottom value is the observed (approximate) value *L*/2ℓ. For example, the first observation of the previous table would be given as:

| g/δ | 5 |
|---|---|
| Δ | **1.88** (!) |
| 2r/δ | 73 |
| L/2ℓ | 6.5 |

---

70 Changed to a superscript in this document.

| $n$ | | $\Delta = L/2\ell$ (calc.) $- L/2\ell$ (obs.) | | | | | | | | | |
|---|---|---|---|---|---|---|---|---|---|---|---|
| 2 | $g/\delta$ | 1.2 | 2 | 2 | 2 | 3.5 | | 5 | 5 | 5 | 11 |
| | $\Delta$ | **1.20** | **1.64** [t] | **1.44** [t] | **1.32** | **1.84** [t] | | **1.79** | **1.88** [t] | **1.69** [t] | **1.96** |
| | $2r/\delta$ | 1.47 | 41 | 55 | 61 | 73 | | 61 | 73 | 135 | 147 |
| | $L/2\ell$ | 9.1 | 6.0 | 6.8 | 7.2 | 6.7 | | 6.2 | 6.4 | 7.9 | 7.2 |
| 3 | $g/\delta$ | 1.2 | | 2.2 | | 3.5 | | 4.5 | 5 | 5 | |
| | $\Delta$ | **1.39** | | **1.37** | | **1.83** [t] | | **1.72** | **1.76** [t] | **1.80** [t] | |
| | $2r/\delta$ | 100 | | 42 | | 73 | | 42 | 73 | 100 | |
| | $L/2\ell$ | 11.5 | | 8.2 | | 8.7 | | 6.8 | 8.1 | 9.1 | |
| 4 | $g/\delta$ | | | 2.2 | | 3.5 | | 4 | 5 | 5 | |
| | $\Delta$ | | | **1.46** | | **1.80** [t] | | **1.81** | **1.75** [t] | **1.78** [t] | |
| | $2r/\delta$ | | | 25 | | 73 | | 25 | 60 | 73 | |
| | $L/2\ell$ | | | 7.8 | | 10.4 | | 6.1 | 8.8 | 9.5 | |
| 5 | $g/\delta$ | | | 2.2 | | 3.5 | | 4 | 5 | 5 | |
| | $\Delta$ | | | **1.52** | | **1.43** [t] | | **1.71** | **1.42** [t] | **1.62** [t] | |
| | $2r/\delta$ | | | 25 | | 73 | | 25 | 58 | 73 | |
| | $L/2\ell$ | | | 8.7 | | 12.2 | | 6.7 | 9.8 | 10.8 | |
| 6 | $g/\delta$ | | | 2.2 | | 3.5 | 3.5 | 4 | 5 | 5 | |
| | $\Delta$ | | | **1.29** | | **1.52** [t] | **1.57** [t] | **1.28** | **1.61** [t] | **1.81** [t] | |
| | $2r/\delta$ | | | 21 | | 48 | 73 | 21 | 46 | 73 | |
| | $L/2\ell$ | | | 8.6 | | 11.0 | 13.4 | 6.5 | 9.2 | 11.5 | |
| 7 | $g/\delta$ | | | 2.2 | | 3.5 | 3.5 | 4 | 5 | 5 | |
| | $\Delta$ | | | **1.31** | | **1.27** [t] | **1.04** [t] | **1.47** | **1.46** [t] | **1.53** [t] | |
| | $2r/\delta$ | | | 21 | | 48 | 73 | 21 | 46 | 73 | |
| | $L/2\ell$ | | | 9.1 | | 12.0 | 15.0 | 6.6 | 9.8 | 12.6 | |
| 8 | $g/\delta$ | | | 2.2 | | 3.5 | | 4 | 5 | 5 | |
| | $\Delta$ | | | **1.15** | | **1.18** [t] | | **1.58** | **1.30** [t] | **1.52** [t] | |
| | $2r/\delta$ | | | 21 | | 48 | | 21 | 46 | 73 | |
| | $L/2\ell$ | | | 7 | | 12.8 | | 6.7 | 10.4 | 13.4 | |
| 9 | $g/\delta$ | | | 2.2 | | 3.5 | | 4 | 5 | 5 | |
| | $\Delta$ | | | **1.20** | | **1.34** [t] | | **1.60** | **1.27** [t] | **1.63** [t] | |
| | $2r/\delta$ | | | 21 | | 48 | | 21 | 46 | 73 | |
| | $L/2\ell$ | | | 10.1 | | 13.2 | | 6.9 | 10.8 | 13.9 | |
| 10 | $g/\delta$ | | | 2.1 | | 3.5 | 3.5 | | 5 | | |
| | $\Delta$ | | | **1.49** | | **1.79** | **1.37** [t] | | **1.23** [t] | | |
| | $2r/\delta$ | | | 21 | | 21 | 48 | | 46 | | |
| | $L/2\ell$ | | | 10.4 | | 7.4 | 13.7 | | 11.1 | | |

[t] [Precisely uniform pitch achieved by machining a helical groove into the coil core].

Besides the already mentioned result that $[\Delta]$[71] slightly decreases as $n$ increases [p601], the table also shows that at constant $n$ (eg., at $n = 2$), $\Delta$ grows with increasing $g/\delta$, but this is less for larger $n$. However, consistent growth of $[\Delta]$ is not to be seen from the table, within the range of $g/\delta = 3.5$ to 5, so that the values of $[\Delta]$ may be combined to mean values for that interval.

There is no clearly recognizable dependence (in a uniform sense) of $\Delta$ on $2r/\delta$ at constant $n$ and constant $g/\delta$. Combining the values of $[\Delta^{(!)}]$ in the range $g/\delta = 3.5$ to 5 to mean values, the following is obtained:

| $n$ | 2 | 3 | 4 | 5 | 6 | 7 | 8 | 9 | 10 |
|---|---|---|---|---|---|---|---|---|---|
| $\Delta^{(!)}$ | 1.80 | 1.78 | 1.78 | 1.49 | 1.63 | 1.33 | 1.33 | 1.41 | 1.30 |

Smoothing these values graphically using a curve, for the case $g/\delta = 1.2$ and $g/\delta = 2$ to 2.2, gave the following result:

Table of $\Delta$ values

| $n$ | $g/\delta$ | | | |
|---|---|---|---|---|
| | 1.2 | 2 | 3.5 to 5 | 11 |
| 2 | 1.3 | 1.54 (!) | 1.80 (!) | 1.96 |
| 3 | 1.29 | 1.5 | 1.78 (!) | — |
| 4 | — | 1.43 | 1.74 (!) | — |
| 5 | — | 1.38 | 1.61 (!) | — |
| 6 | — | 1.32 | 1.50 (!) | — |
| 7 | — | 1.29 | 1.40 (!) | — |
| 8 | — | 1.28 | 1.35 (!) | — |
| 9 | — | 1.27 | 1.32 (!) | — |
| 10 | — | 1.26 | 1.30 (!) | — |

The values of $\Delta$ for $g/\delta = 3.5$ to 5 are more reliable than the others. For Tesla coils, this range of $g/\delta$ is the most important.

***The self-inductance of a coil of not more than* 10 *turns* (capacitor excluded**[72]**) *is represented by the formula***[73] :

(35) $\quad \dfrac{L}{2\ell} = n\left\{1 + \dfrac{h^2}{32r^2}\log_e\left(\dfrac{8r}{\sqrt{h^2+\delta^2}}\right) - y_1 + \dfrac{h^2}{16r^2}y_2\right\} + \log_e\left(\dfrac{g}{\delta}\right) - \Delta \quad ,$

---

71 Drude puts $B$ here, but clearly means $\Delta$. It is probable that he used $B$ originally, but realised that it had also been used for the transmission-line shorting bügel (fig. 3), and so changed to $\Delta$ but missed a few instances.

72 The word used here was 'entladungen', strictly 'discharges' but this is to be interpreted in the non-electrical sense; 'dismissed' or 'taken away'.

73 [604-1] In this formula, the term $\frac{1}{3}\delta^2 : 32\ r^2$ is negligible compared to 1 because the formula achieves a claimed accuracy of only about 1%, and $g/\delta$ in practice is always much less than 1.

where $y_1$ and $y_2$ are from the following (Stefan's) table:

| δ/h | $y_1$ | $y_2$ | δ/h | $y_1$ | $y_2$ |
|---|---|---|---|---|---|
| 0.00 | 0.500 | 0.13 | 0.55 | 0.808 | 0.34 |
| 0.05 | 0.549 | 0.13 | 0.60 | 0.818 | 0.38 |
| 0.10 | 0.592 | 0.13 | 0.65 | 0.826 | 0.43 |
| 0.15 | 0.631 | 0.14 | 0.70 | 0.833 | 0.47 |
| 0.20 | 0.665 | 0.15 | 0.75 | 0.838 | 0.52 |
| 0.25 | 0.695 | 0.17 | 0.80 | 0.842 | 0.58 |
| 0.30 | 0.722 | 0.19 | 0.85 | 0.845 | 0.63 |
| 0.35 | 0.745 | 0.22 | 0.90 | 0.847 | 0.69 |
| 0.40 | 0.765 | 0.24 | 0.95 | 0.848 | 0.75 |
| 0.45 | 0.782 | 0.27 | 1.00 | 0.848 | 0.82 |
| 0.50 | 0.796 | 0.31 | | | |

(For $n = 1$, $h = 0$, $g/\delta = 1$, i.e., for simple circular loops, equation (35) simplifies to equation (28), and in this case $\Delta = 0.81$ [74] is used.)

Comparing the observed [1] [uniform pitch] values of $L/2\ell$ from the table on page 602, to the calculated values from equation (35) with smoothed $\Delta$ values from the table on page 603, there are 11 cases where the values deviated by more than 1% (but not more than 2.5%), and 19 cases where the values are in agreement to within 1%. Formula (35) is therefore accurate to 1% for the range $g/\delta = 3.5$ to 5, and 2% for other $g/\delta$ values.

---

74 [604-2] This value is not in contradiction with the Table of $\Delta$, since $\Delta$ decreases sharply as $g/\delta$ decreases to 1.

## 20. Rigorous testing and application of the formulae in two Tesla transformers

**a)** The secondary winding of a Tesla transformer, consisting of 268 turns of 1 mm thick copper wire, was wound on a hollow ebonite cylinder of 8 mm wall thickness and 6.4 cm outer diameter. The coil height was h = 43 cm, the wire length $\ell$ = 5480 cm, and the pitch g = 1.6 mm. Since the coil diameter is equal to 6.4 + 0.1 = 6.5 cm, then $h$:$2r$ = 6.6. At this ratio of $h/2r$ and the value $g/\delta$ = 1.6, the parameter $f$, in accordance with the table on p322, would be 0.66 for a hollow core, 0.71 for a solid core. Because its wall thickness is ⅛$^{th}$ of the coil diameter, this coil core more closely approaches the solid ebonite core; so it can be assumed approximately that $f$ = 0.70. Therefore, the self-resonant half-wavelength of the secondary winding is:

$$\lambda/2 = \ell \cdot f = 3840 \text{ cm}$$

The primary coil consisted of five turns of 1.4 mm thick wire and had the following constants:

$$n = 5, \ h = 4.5 \text{ cm}, \ 2r = 12.4 \text{ cm}, \ \delta = 1.4 \text{ mm}, \ g = 1.1 \text{ cm}, \ \ell = 195 \text{ cm}.$$

Since $\delta/h$ = 0.031, in equation (35) $y_1$ and $y_2$ have the values: $y_1$ = 0.53, $y_2$ = 0.13 (according to the table on [p604] ). Additionally, $g/\delta$ = 7.85 and $n$ = 5, so in formula (35), the value of $\Delta$ from the table on p602 will be about 1.67. (The uncertainty of 5%, i.e., the assumption $\Delta$ = 1.72, makes only 0.5% error in the calculation of the self-inductance of $L$). Therefore, from (35)

$$\frac{L}{2\ell} = 9.94 \ , \text{ i.e., } L = 3875 \text{ cm [ 3.875 µH ]}.$$

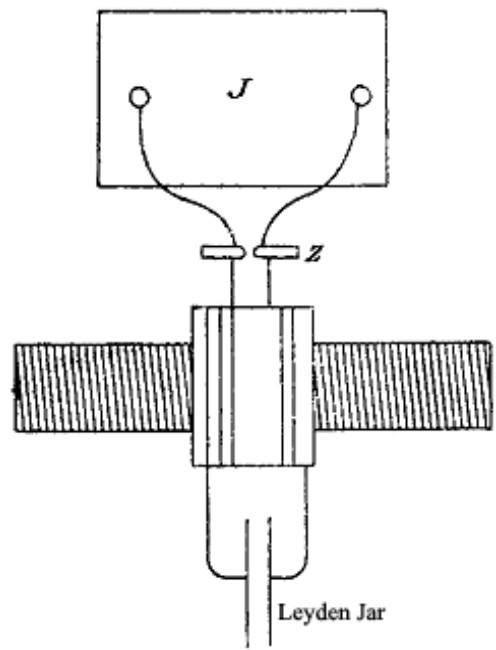

Leyden Jar

This coil was now opened at two facing points[75] [and connected] to four straight wires, two of which led to the zinc spark gap (length of the wires $\ell'$ = 7 cm, mean distance $d'$ = 5 cm, wire thickness $2\rho'$ = $\delta$ = 1.4 mm), while the two others led to a small Leyden jar ($\ell'$ = 9 cm, $d'$ = 10 cm, $2\rho'$ = $\delta$ = 1.4 mm) [see illustration]. Therefore according to equation (29) of p596 for $L$, the two values: $L'$ = 120 cm and $L'$ = 179 cm are added, so that the entire self-inductance $L$ of the primary circuit has a value of $L_1$ = 4174 cm.

The Leyden jar had an inner [conductive covering] height of 10.2 cm, 6 cm inner diameter, and 2.6 cm glass thickness. The area of the inner tinfoil covering was therefore
$S = \pi (6 \times 10.2 + 3^2) = 70\pi \text{ cm}^2$,
and
$S:4\pi d = 70:1.04 = 67.5 \text{ cm}.$

---

75 [605-1] The circuit is accurately described by me in **'Zur Messung der Dielektricitätsconstante vermittelst elektrischer Drahtwellen'** [Measuring the dielectric const. by means of electric wire waves (standing waves)], Ann. der Phys. 313(6) (4$^{th}$ series vol. 8). p336-347. 1902.

The Leyden jar was chosen so that the transformer worked well, i.e., resonance between the primary circuit and secondary circuit was approximately[76] established. To produce the half-wavelength $\lambda/2 = 3840$ cm on the secondary winding with a self-inductance $L = 4174$, then according to formula (27′) page 593 the capacitance is calculated to be $C = 358$ cm. Since for the Leyden jar $C = \varepsilon.S : 4\pi d$, where $\varepsilon$ is the dielectric constant of the glass, $\varepsilon$ was calculated as:

$$\varepsilon = 358:67.5 = 5.31 \,.$$

In fact, Löwe[77], using fast electrical oscillations, has observed dielectric constants between 5 and 7.7, depending on the type of glass.

Thus, this calculation shows agreement, as far as one can expect, especially since the primary circuit tuning was taken quite crudely[78] from the secondary circuit, and (see footnote [606-1]) probably the primary circuit had a somewhat larger half-wavelength than $\lambda/2 = 3840$, hence $C$ was perhaps a little larger than 358 cm and $\varepsilon$ slightly greater than 5.31.

**b)** A second smaller Tesla transformer, with a secondary winding of 127 turns of 1 mm thick copper wire, was also wound on a hollow ebonite cylinder of 8 mm wall thickness.

- [p607] -

From $h = 24.8$ cm, $2r = 6.5$ cm, i.e., $h/2r = 4$, $g/\delta = 2$, and, according to the table on p322, $f = 0.81$, there results

$$\lambda/2 = \ell \cdot f = 2590 \cdot 0.81 = 2100 \text{ cm}.$$

The primary coil had the constants:

$n = 3$,  $h = 4.5$ cm,  $2r = 12.7$ cm,  $\delta = 1.4$ mm,  $g = 2.2$ cm,  $\ell = 120$ cm.

Therefore, according to equation (35) ($\Delta$ is assumed to be 1.92, as $g/\delta = 16$) $L = 1590$ cm. The leads to the coil ($\ell' = 7$ cm, $d' = 6$ cm and 8 cm respectively) give $L' = 124$ cm and $L' = 132$ cm respectively, so that $L_1 = 1846$ cm. To resonate with the secondary circuit, the capacitance of the Leyden jar was therefore $C = 242$ cm. The determined values were: height, 6.8 cm; inner diameter, 5.5 cm; glass thickness, 2.6 mm; therefore S: $4\pi d = 43.2$ cm. Hence:

$$\varepsilon = 242: 43.2 = 5.60 \,.$$

The natural oscillation period of the primary circuit was now measured directly using the method mentioned on p598 [79]; [i.e.,] by removing the secondary coil of the transformer and feeding the zinc spark gap of the primary from an induction coil or from another Tesla transformer, and then making the oscillations generated in the primary circuit act, with weak inductive coupling, on the secondary circuit, described on p598, [i.e.,] a [transmission] line of two 1 mm thick, parallel wires of 2.65 cm distance, over which an adjustable metal shorting-strap *B* could be moved.  At the other end, the parallel wires were bent at right angles and were, using their slightly springy nature, applied to the plates of a circular-plate capacitor of 12.1 cm diameter and 1 mm plate thickness, the plates being separated by three ebonite plates of (on average) 0.53 mm thickness and 9 mm$^2$ area.  The capacitance *C* of this capacitor is calculated according to the formula on p299 to be *C* = 177.6 cm [197.6 pF].  A Zehnder's vacuum tube[80] was applied to a plate on this capacitor, and the bracket *B* moved by hand along the parallel wires.

---

76 [606-1] A Tesla transformer shows the strongest effect when the primary circuit has a slightly smaller natural period than the secondary circuit, as I intend to demonstrate in a later article. However, this deviation of the natural period of both circuits is not significant, and only noticeable for very strong coupling, i.e., the primary circuit very close to the secondary circuit. In the present transformer, this deviation is estimated to be less than 5%.

77 [606-2] '**Experimentel-untersuchung über electrische dispersion einiger organiscuer Säuren, Ester, und von zehn Glassorten**' [Experimental study of electrical dispersion of some organic acids, esters, and 10 types of glass.], K. F. Löwe, Ann. Phys. 302 (11) 1898 (Wied. Ann. 66), p390-410, 582-596. 1898. See p402.

78 [606-3] For the production of Tesla transformers, the stronger the magnetic coupling between the primary and secondary circuit, the less exact this tuning needs to be.

79 [607-1] This method is thus also suitable for long waves, such as those produced by Tesla transformers, and is very convenient for determination of the frequency of the oscillations.

- [p608] -

In a fairly[81] sharply determinable resonance position of *B*, the vacuum tube glowed brightly.  The length of the rectangular secondary line was then  *a* = 157 cm.  The total length of the secondary line was  $\ell$ = 2.157 + 2.65 + 2.35 = 319 cm.  Therefore the self-inductance, using equation (33) on p597, is *L* = 2520 cm.  The self-resonance half-wavelength of the secondary line, i.e., also of the primary line, is therefore according to equation (26) on p593,  $\lambda/2$ = 2120 cm.  So we arrive at a value for $\lambda/2$ that matches very well with the self-resonance half-wavelength of the secondary coil of the Tesla transformer as calculated on p607.  — Based on the value  $\lambda/2$ = 2120 cm, the capacitance of the Leyden jar in the primary circuit of the Tesla transformer is calculated to be 246 cm, that is, the dielectric constant ε of the glass of the Leyden jar is

$\varepsilon$ = 246:43.2 = 5.7 .

---

80 '**Zur objectiven Darstellung der Hertz'schen Versuche über Strahlen electrischer Kraft'**,  L. Zehnder, Ann. Phys. 283(9) (Wied. Ann. 47) 1892, p77-92, see p82 ; see also '**Objective representation of Hertz's researches on electrical radiation**' (abstract of the above), L Zehnder, The Electrician, Dec. 30, 1892, vol. 30, p258.
 — In addition to the normal pair of electrodes, a Zehnder gas-discharge tube has a pair of electrodes that can be used for triggering or priming.  A Zehnder tube is normally provided with DC bias, across one pair of electrodes, sufficient to make it strike in the presence of a triggering signal or some other electromagnetic disturbance.
81 [608-1]  The resonance position is not determined as sharply as when using a petroleum or air capacitor, because of emission of corona discharge.  Perhaps something disturbs electrical absorption in the glass of the Leyden jar.  Löwe (as cited above) was however not able to detect electrical absorption in the glass for much faster vibrations.

## Summary of results

     **1**. *The natural period of a coil increases with the dielectric constant of the coil core and its surroundings* (e.g., transformer immersed in oil).

     **2**.  The dielectric constant of ebonite for Hertzian oscillations is  $\varepsilon = 2.79$.  Ebonite is electrically isotropic.

     **3**.  If the pitch of the turns at the middle of a coil is smaller than that of the end turns, the natural oscillation period of the coil is slower than in the opposite case or at constant pitch.

<div align="center">- [p609] -</div>

     **4**.  *The self-resonance half-wavelength $\lambda/2$ for a constant coil pitch g depends on the coil wire length $\ell$, the coil height h, the coil diameter 2r, the wire thickness $\delta$ such that*

$$\lambda/2 = \ell\, f(h/2r\,,\ g/\delta\,,\ \varepsilon)\,,$$

*where $\varepsilon$ is the dielectric constant of the coil core*.  On p322 - 323 tables of $f$ for practically occurring cases (for hollow core coils and air-core coils) are given.
    *Within certain limits $f\sqrt{h/r}$ is constant*.  On p329, the numerical values of this product are specified,  hence $\lambda/2$ can also be easily calculated.

     **5**.  *The self-resonance half-wavelength of nearly closed circular loops of thin wire is* 6.5% *larger than the wire length*.

     **6**.  *The overtone resonances of a coil are not harmonically related to the fundamental resonance*, and the ratio of the frequencies of the fundamental and overtone oscillations is somewhat dependent on the ratio *h/2r.*  With the subsequent possible overtones, it is found that (with decreasing intensity), the first overtone produces two current anti-nodes in the coil, the second three current anti-nodes, etc.  In the overtone resonances, the coil does not oscillate in congruent parts.

     **7**.  *By applying a capacitance to the free end of a coil, the natural period of the fundamental oscillation of a coil is increased in a calculable way* (p337) *and experimentally confirmed.  This increase is always smaller than twice the period of the coil with free ends*.

     **8**.  *On p604 a formula is given for calculating the self-inductance of short coils with fast alternating currents*.

     **9**.  *This formula, and the tables on  p322, 323 or p329, give the ability to calculate the correct primary circuit capacitance for every Tesla transformer*.  If the Tesla transformer secondary coil is not free-ended, but is rather connected to one or two capacitances, then the best capacitance to select for the primary circuit is the greater one, and this may also may be calculated in advance from the preceding data.  Two samples calculations for two different Tesla transformers have confirmed the applicability of the formula and the tables.

— Grounding of one end of the Tesla transformer secondary winding gives no definite result[82].

**10**. ***Secondary windings on wood or cardboard tubes are not as good*** (due to electrical absorption in wood or cardboard) ***as coils on ebonite, or glass, or coreless coils***. Therefore, for effective construction of Tesla transformers, the former are less favourable than the latter. — Also, for the capacitance of the primary circuit of the Tesla transformer, it is better to use metal plates in a petroleum bath than a Leyden jar, because of the corona discharges on the tinfoil lining (and perhaps also due to electrical absorption of the glass). — It is advisable to make the primary circuit with a small number (1 - 3) turns of thick wire (2 - 4 mm) (so that the self-inductance is as small as possible), while it is advisable to make the secondary circuit from thinner wire (0.5 mm) wound into a coil whose height about twice its diameter[83].

**11**. ***By using a liquid-immersed circular plate capacitor*** of 10 cm radius, which is made to oscillate with a circular loop of 21 cm in diameter and 3 mm thickness, the wavelength can be changed continuously — by varying the plate distance of the capacitor, and the liquid of its bath — over a large interval from 3 m (5 cm plate distance, air between capacitor plates) up to 133 m (1 mm plate distance, water between capacitor plates).

With water filling, a Tesla transformer must be used to initiate the excitations (this is convenient in any case).

Giessen, June 1902.

(Received 26th June 1902)

---

82 [610-1]  See F. Braun, Ann. d. Phys. 8. p209. 1902.
83 [610-2]  I have considered the theoretical aspects of these results, which I intend to publish later.

## Appendix 1.  Recalculation of the table on p328.
(Added by the translators).

Values of $f = \lambda / 2\ell$ , for coreless coils.

| h/2r | h/r | r/h | p322 table g/δ = 1.09 f | p327 eq(9) f / √α | √α | α | p328 table 2 √(α π) |
|---|---|---|---|---|---|---|---|
| 0.04 | 0.08 | 12.50 | 2.970 | 7.206 | 0.412 | 0.170 | 1.461 |
| 0.05 | 0.10 | 10.00 | 2.980 | 6.430 | 0.463 | 0.215 | 1.643 |
| 0.06 | 0.12 | 8.33 | 3.000 | 5.854 | 0.513 | 0.263 | 1.817 |
| 0.07 | 0.14 | 7.14 | 2.960 | 5.402 | 0.548 | 0.300 | 1.942 |
| 0.08 | 0.16 | 6.25 | 2.930 | 5.035 | 0.582 | 0.339 | 2.063 |
| 0.09 | 0.18 | 5.56 | 2.880 | 4.728 | 0.609 | 0.371 | 2.159 |
| 0.10 | 0.20 | 5.00 | 2.790 | 4.467 | 0.625 | 0.390 | 2.214 |
| 0.15 | 0.30 | 3.33 | 2.520 | 3.557 | 0.708 | 0.502 | 2.511 |
| 0.20 | 0.40 | 2.50 | 2.320 | 2.998 | 0.774 | 0.599 | 2.743 |
| 0.25 | 0.50 | 2.00 | 2.180 | 2.613 | 0.834 | 0.696 | 2.957 |
| 0.30 | 0.60 | 1.67 | 2.080 | 2.333 | 0.892 | 0.795 | 3.161 |
| 0.35 | 0.70 | 1.43 | 2.000 | 2.121 | 0.943 | 0.889 | 3.343 |
| 0.40 | 0.80 | 1.25 | 1.925 | 1.956 | 0.984 | 0.968 | 3.488 |
| 0.50 | 1.00 | 1.00 | 1.790 | 1.720 | 1.041 | 1.084 | 3.690 |
| 0.60 | 1.20 | 0.83 | 1.670 | 1.558 | 1.072 | 1.149 | 3.800 |
| 0.70 | 1.40 | 0.71 | 1.560 | 1.439 | 1.084 | 1.175 | 3.842 |
| 0.80 | 1.60 | 0.63 | 1.470 | 1.348 | 1.091 | 1.190 | 3.867 |
| 0.90 | 1.80 | 0.56 | 1.390 | 1.273 | 1.092 | 1.191 | 3.869 |
| 1.00 | 2.00 | 0.50 | 1.330 | 1.212 | 1.098 | 1.205 | 3.891 |
| 1.20 | 2.40 | 0.42 | 1.225 | 1.113 | 1.101 | 1.212 | 3.903 |
| 1.40 | 2.80 | 0.36 | 1.150 | 1.035 | 1.111 | 1.234 | 3.938 |
| 1.60 | 3.20 | 0.31 | 1.090 | 0.972 | 1.121 | 1.257 | 3.974 |
| 1.80 | 3.60 | 0.28 | 1.040 | 0.920 | 1.131 | 1.279 | 4.009 |
| 2.00 | 4.00 | 0.25 | 0.995 | 0.875 | 1.138 | 1.294 | 4.033 |
| 2.20 | 4.40 | 0.23 | 0.960 | 0.836 | 1.149 | 1.320 | 4.073 |
| 2.40 | 4.80 | 0.21 | 0.925 | 0.801 | 1.154 | 1.333 | 4.092 |
| 2.60 | 5.20 | 0.19 | 0.900 | 0.771 | 1.168 | 1.363 | 4.139 |
| 2.80 | 5.60 | 0.18 | 0.875 | 0.744 | 1.177 | 1.385 | 4.172 |
| 3.00 | 6.00 | 0.17 | 0.845 | 0.719 | 1.175 | 1.381 | 4.166 |
| 3.50 | 7.00 | 0.14 | 0.795 | 0.667 | 1.192 | 1.422 | 4.227 |
| 4.00 | 8.00 | 0.13 | 0.760 | 0.624 | 1.217 | 1.482 | 4.315 |
| 4.50 | 9.00 | 0.11 | 0.735 | 0.589 | 1.248 | 1.557 | 4.423 |
| 5.00 | 10.00 | 0.10 | 0.715 | 0.559 | 1.279 | 1.635 | 4.533 |
| 5.50 | 11.00 | 0.09 | 0.700 | 0.533 | 1.312 | 1.723 | 4.653 |
| 6.00 | 12.00 | 0.08 | 0.685 | 0.511 | 1.341 | 1.798 | 4.754 |

## Appendix 2.  Obscure or obsolete German words and abbreviations.
§ = section
aichen = eichen = calibrate
aichung = eichung = calibrated
aussprühenden = emission/emitting
axe = acshe = axis (eg., Axenverhältnis, Draht-axe, Spulen-axe, Axenabstand)
bauch / bäuche = bump, maximum, anti-node
bewickelung = bewicklung = winding
beob. = beobachtung = observation (obs.)
ber. = berechnung = calculation (calc.)
bez. = bezugnahme = with reference to, compared to (ref.)
bez. = bezüglich = regarding
bez. = beziehungsweise = respectively, or
Büschelentladungen = corona discharge
Capacität = Kapazität = capacity (i.e., capacitance)
Columnen = Kolumnen = columns
constatirt = konstatiert = verified, specified, fixed, stated, agreed
correcturbedürftig = in need of correction
definirtes =  definiertes = defined
deformirbar = deformierbar = deformable
Dielektricitätsconstante = Dielektrizitätskonstante
Drahtleitung = wire circuit
Eigenwellenlänge = inherent or characteristic wavelength (i.e., self-resonance wavelength)
ergiebt = ergibt = yields, results in, comes to, etc.
i/B. = i. B. = im Besonderen = specifically, in particular
inductorium = induktorium = induction coil
isolierend = isolirend = insulating
Kugelfunctionen = spherical harmonics (Kugelfunktionen = spherical functions)
l. c. = loc. cit. = loco citato = in the place already cited (i.e., in the same place in an earlier reference).
lognat() = $\log_e()$ = ln() = natural logarithm
multiplicren = multiplizieren = multiply
neubewickelung = neubewicklung = rewinding
niedrigen = low, squat, short
Proc. = procentische = Prozentsatz = percentage = per cent, %
reducirten = reduzierten = reduce, decrease
Schliessungsdraht = wire loop
Schliessungskreises = closed loop or closed circuit
Schwingung = oscillation or vibration
tg( ) = tan( ) = tangent
Vergl. = vergleichen = compare
Zwirnsfäden (lit. twisted threads) = Bindfaden = twine, string

**Note**:
c  is replaced by  k  or  z  in modern spellings, Hence: Dielektricitätsconstante = Dielektrizitätskonstante.
Other spelling changes result from dropping or insertion of vowels; e.g., ergiebt = ergibt, etc..